\documentclass[twocolumn,showpacs,floats,floatfix,superscriptaddress,aps,prl,longbibliography]{revtex4-2}
\usepackage{amsfonts,amssymb,amsmath}
\usepackage{mathtools}
\usepackage{mathrsfs}
\usepackage{color,calc}
\usepackage{graphicx}
\usepackage{epstopdf}
\usepackage{bm}
\usepackage[normalem]{ulem}
\usepackage[colorlinks=true, linkcolor=blue, citecolor=blue,urlcolor=blue]{hyperref}

\def\i{\,\text{i}}

\def\i{i}

\def\to{\rightarrow}

\def\to{\rightarrow}

\def\i{\text{i}}

\usepackage[normalem]{ulem}

\begin{document}

\author{Qingyun Cao}
\thanks{These authors contributed equally.\\
\href{qingyun.cao@uni-ulm.de}{qingyun.cao@uni-ulm.de}\\
\href{yaomingchu@hust.edu.cn}{yaomingchu@hust.edu.cn}\\
\href{jianmingcai@hust.edu.cn}{jianmingcai@hust.edu.cn}\\
\href{yu.liu@uni-ulm.de}{yu.liu@uni-ulm.de}}
\affiliation{Institute for Quantum Optics, Ulm University, Albert-Einstein-Allee 11, Ulm 89081, Germany}

\author{Genko T. Genov}
\thanks{These authors contributed equally.\\
\href{qingyun.cao@uni-ulm.de}{qingyun.cao@uni-ulm.de}\\
\href{yaomingchu@hust.edu.cn}{yaomingchu@hust.edu.cn}\\
\href{jianmingcai@hust.edu.cn}{jianmingcai@hust.edu.cn}\\
\href{yu.liu@uni-ulm.de}{yu.liu@uni-ulm.de}}
\affiliation{Institute for Quantum Optics, Ulm University, Albert-Einstein-Allee 11, Ulm 89081, Germany}

\author{Yaoming Chu}
\affiliation{School of Physics, Center for Intelligence and Quantum Science (CIQS), Hubei Key Laboratory of Gravitation and Quantum Physics, International Joint Laboratory on Quantum Sensing and Quantum Metrology, Huazhong University of Science and Technology, Wuhan 430074, China}

\author{Jianming Cai}
\affiliation{School of Physics, Center for Intelligence and Quantum Science (CIQS), Hubei Key Laboratory of Gravitation and Quantum Physics, International Joint Laboratory on Quantum Sensing and Quantum Metrology, Huazhong University of Science and Technology, Wuhan 430074, China}

\author{Yu Liu}
\affiliation{Institute of Theoretical Physics and IQST,  Ulm University, Albert-Einstein-Allee 11, Ulm 89081, Germany}

\author{Alex Retzker}
\affiliation{Racah Institute of Physics, The Hebrew University of Jerusalem, Jerusalem 91904, Givat Ram, Israel}
\affiliation{AWS Center for Quantum Computing, Pasadena, California 91125, USA}

\author{Fedor Jelezko}
\affiliation{Institute for Quantum Optics, Ulm University, Albert-Einstein-Allee 11, Ulm 89081, Germany}

\title{Overcoming frequency resolution limits using a solid-state spin
quantum sensor}

\date{\today}

\begin{abstract}
The ability to determine precisely the separation of two frequencies is fundamental to spectroscopy, yet the resolution limit poses a critical challenge: distinguishing two incoherent signals becomes impossible when their frequencies are sufficiently close. Here, we demonstrate a simple and powerful approach, dubbed {\it superresolution quantum sensing}, which experimentally resolves two nearly identical incoherent signals using a solid-state spin quantum sensor. By carefully choosing interrogation times that satisfy the superresolution condition, we eliminate quantum projection noise, overcoming the vanishing distinguishability of signals with near-identical frequencies. This leads to improved resolution, which scales as $t^{-2}$ in comparison to the standard $t^{-1}$ scaling. Together with a greatly reduced classical readout noise assisted by a nuclear spin, we are able to achieve sub-kHz resolution with a signal detection time of 80 microseconds. Our results highlight the potential of quantum sensing to overcome conventional frequency resolution limitations, with broad implications for precision measurements. 
\end{abstract}

\maketitle

Quantum metrology and quantum sensing employ the fundamental laws of quantum physics to achieve precision that is typically not feasible with classical analogs \cite{Degen2017Quantum,Gardner2025prxq}. 
For example, reaching the fundamental 
quantum Cram\'er-Rao bound has allowed for resolution of two weak thermal optical point sources with a distance below the diffraction limit \cite{tsang2016quantum,nair2016far,lupo2016ultimate,chrostowski2017super}. 
The ability to nullify measurement projection noise by coherent control and a suitable measurement basis provides a critical resource for solving the frequency resolution problem: when the separation between 
frequencies vanishes, the uncertainty to distinguish them becomes divergent. Minimizing quantum projection noise mitigates this effect and leads to a finite uncertainty, 
allowing to overcome resolution limits with quantum sensing of noisy signals \cite{gefen2019overcoming,Mouradian2021pra}.

Frequency resolution problems are crucial
in science and technology. As a prominent example, in nanoscale nuclear magnetic resonance (NMR), dominated by statistical nuclear polarization, the spin precession in an external magnetic field typically generates oscillating magnetic signals with short coherence times \cite{mamin2013nanoscale,staudacher2013nuclear,Staudenmaier2023Optimal}. This poses severe challenges in determining complex spectra with quantum sensors \cite{rotem2019limits,Mouradian2021pra}, even with advanced quantum sensing techniques, such as correlation spectroscopy \cite{Laraoui2013High,Staudacher2015Probing,Kong2015Towards,Pfender2017Nonvolatile,Rosskopf2017Quantum,aslam2017nanoscale,Pfender2019High,Cujia2019Tracking} and quantum heterodyne detection \cite{schmitt2017submillihertz,Boss2017Quantum,Glenn2018High,Chu2021PRapplied}.
%

\begin{figure}[bht]
\centering
\includegraphics[width=1\linewidth]{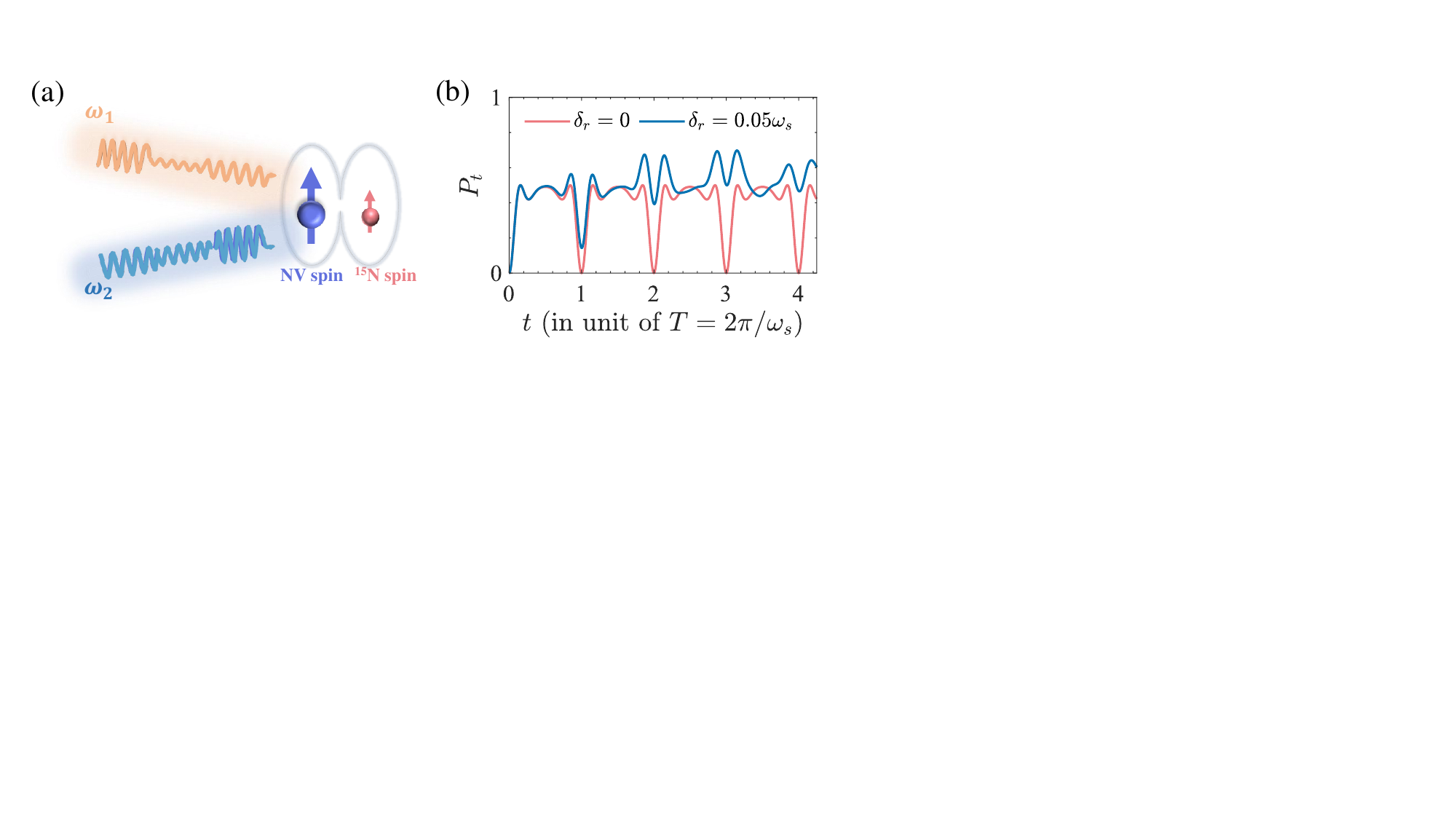}
\caption{ Principles of superresolution quantum sensing. (a) Single NV spin as a quantum sensor (with $\prescript{15}{}{\mathbf{N}}$ nuclear spin serving as a memory qubit) for detecting two incoherent oscillating signals with the frequencies $\{\omega_1,\omega_2\}$ (cf. Eq.\eqref{Eq:Hs-Main}). (b) The transition probability, $P_t$, displays maximal difference for $\delta_{r}\equiv(\omega_1-\omega_2)/2=0$ (red curve) and $\delta_{r}=0.05\omega_{s}$ (blue curve), where $\omega_s=(\omega_1+\omega_2)/2$, in the neighboring region of $\omega_s t=2n\pi$ with $n$ positive integers, as indicated by the superresolution condition. 
}
\label{Principles_super}
\end{figure}
%

\begin{figure*}[t]
\centering
\includegraphics[width=0.9\linewidth]{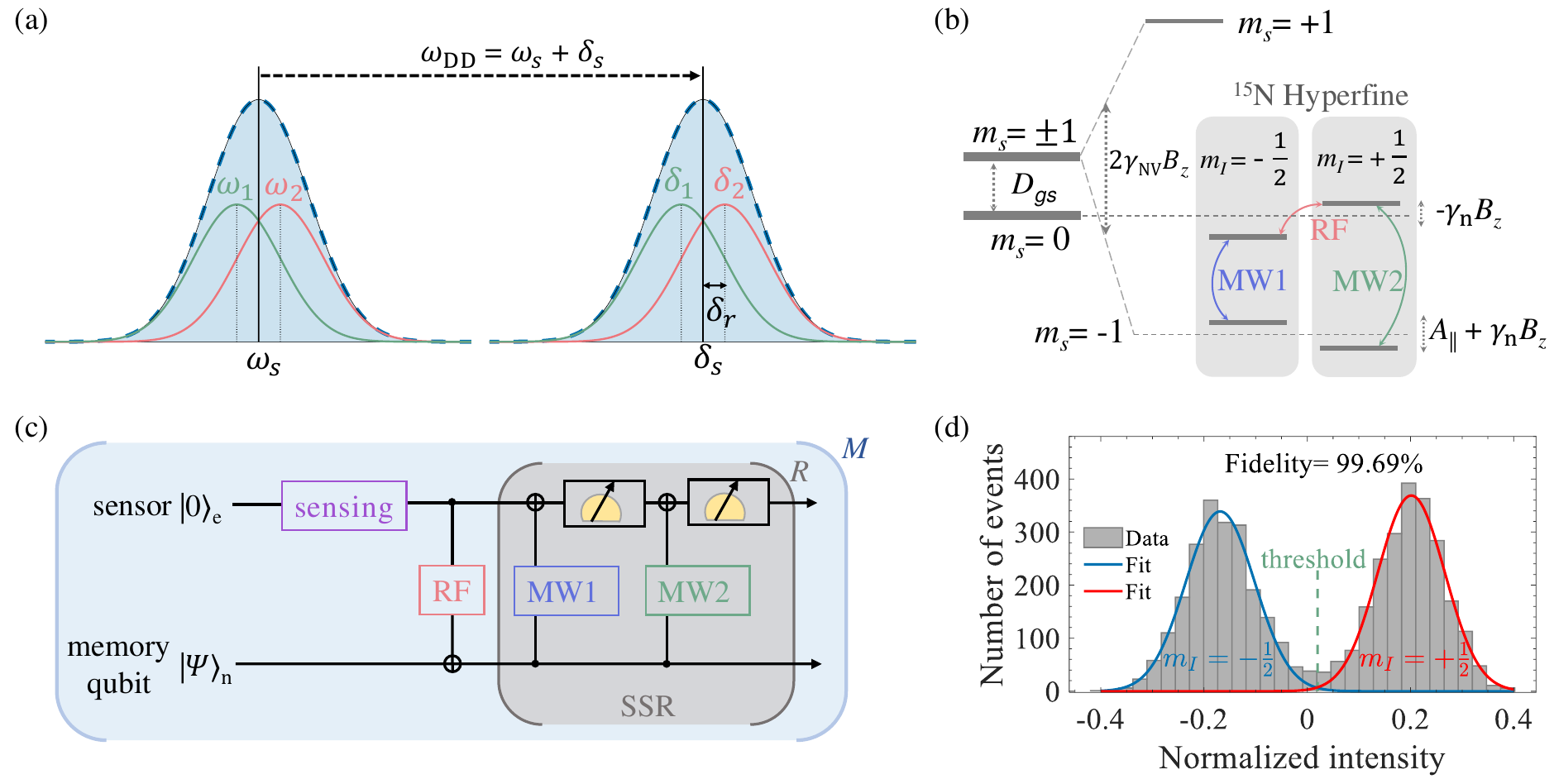}
\caption{ Experimental protocol using a solid-state spin quantum sensor in diamond. (a) Resolving two angular frequencies, $\omega_1$ and $\omega_2$, is challenging as they approach each other. We shift  $\omega_{i}\to\delta_{i}$ in the interaction basis by applying dynamical decoupling \cite{Degen2017Quantum} with short $\pi$ pulses, separated by time $\pi/\omega_{\text{DD}}$. (b) Energy levels of the NV center electron spin with hyperfine coupling to its intrinsic $\prescript{15}{}{\mathbf{N}}$ nuclear spin. The $m_{S}= \pm1$ energy level degeneracy is lifted by a Zeeman splitting $2\gamma B_{z}$ with an external magnetic field $B_{z}$, aligned with the NV center symmetry axis. The hyperfine coupling further splits each electronic state into two sublevels, $m_{I}= \pm1/2$. The nuclear Zeeman splitting, $\gamma_{\text{n}}B_{z}$, is also taken into account due to the high magnetic field. (c) Sensing and readout sequence. The spin state of the sensor is mapped onto the nuclear spin with a controlled RF $\pi$ pulse, which can be repetitively read out by single shot readout (SSR) measurements. (d) The histogram of the fluorescence time trace obtained from SSR readout, which can be well fitted by two Gaussian distributions. The high and low counts correspond to a nuclear spin state $|\uparrow\rangle$, and $|\downarrow\rangle$, respectively, giving rise to a high readout fidelity of 99.69$\%$. }
\label{setup}
\end{figure*}

%
Here, we use a solid-state quantum sensor of a single NV center in diamond to overcome the frequency resolution limit. By meticulously choosing the interrogation time to satisfy a superresolution condition \cite{gefen2019overcoming}, we simultaneously nullify the quantum projection noise and maximize its response to frequency separation. We also improve the readout technique of the solid-state spin sensor by using an ancilla nuclear memory spin \cite{Pfender2017Nonvolatile,Rosskopf2017Quantum,aslam2017nanoscale,lovchinsky2016nuclear,zhao2023sub,meinel2023high}, reducing classical readout noise and improving resolution. We achieve discrimination of two incoherent time-oscillating signals with a frequency separation below the Fourier limit down to the sub-kHz regime, using a solid-state spin with a single measurement detection time of only 80 $\mu$s. We demonstrate the method in the context of pulsed dynamical decoupling and Ramsey sequences, allowing for detection of frequencies from dc to several tens of MHz \cite{Degen2017Quantum}. Our approach resolves a fundamental limitation in spectroscopy, paving the way for high-resolution nanoscale NMR.

\emph{Principles of superresolution quantum sensing.---}
Spectral resolution problems typically involve a time oscillating signal, interacting with a probe, see Fig.\,\ref{Principles_super}(a). By extracting information, encoded in the probe, one can estimate the signal frequencies, e.g., $\omega_1$ and $\omega_2$. Resolution becomes challenging when $\omega_1 \rightarrow \omega_2$, as the uncertainty of the frequency separation estimate $\Delta (\omega_1-\omega_2) \rightarrow \infty$. As an explicit example, we consider nano NMR sensing where the effect of the signal on the qubit sensor is characterized by the Hamiltonian  \cite{staudacher2013nuclear},
\begin{equation}
H_s=\sum_{i=1,2}\Omega_i \sin[\omega_{i}t+\varphi_i (t)]\sigma_z.
\label{Eq:Hs-Main}
\end{equation}
where $\Omega_i$ are the signal amplitudes at the two frequencies, and $\varphi_i(t)$ are their phases, which can vary in time. It is useful to define the angular frequency difference $\delta_r=(\omega_1-\omega_2)/2$ and mean angular frequency $\omega_s=(\omega_1+\omega_2)/2$. We emphasize that we consider the general scenario of incoherent time-oscillating signals where the phases $\varphi_i (t)$ are uncorrelated between different measurements. 

We consider standard Ramsey interferometry where we initialize the sensor qubit in a ground state $|0\rangle$, prepare a coherent superposition state with a $\pi/2$ pulse, and let it evolve for time $t$ under the Hamiltonian in Eq.\,\eqref{Eq:Hs-Main}. We focus on the particular case when $\Omega_{1,2}=\Omega$ to correspond with our experimental implementation but the method also works when this condition is not satisfied \cite{gefen2019overcoming}. The transition probability in the initialization basis is \cite{gefen2019overcoming}
\begin{equation}
\label{Eq:Population}
P_t=\frac{1}{2}\left[1-\prod_{i=1}^2 J_{0} \left(\frac{4\Omega}{\omega_i}\sin\left(\frac{\omega_i t}{2}\right)\right)\right],
\end{equation}
where $J_0(.)$ is the zeroth order Bessel function. As $\delta_r \to 0$ we find that $\partial P_t/\partial\delta_r=0$ \cite{Supplemental}, leading to the aforementioned resolution problem. The vanishing susceptibility arises from the symmetry of interchanging $\omega_1$ and $\omega_2$ in the case of two closely spaced signals, as shown in Eq.\,\eqref{Eq:Hs-Main}.

For small values of $\delta_r$, we can expand the transition probability to its second order, i.e. $P_t\approx a_t+b_t\,\delta_r^2$ with $a_t$ and $b_t$ functions of $\{\Omega_,\omega_s,t\}$ \cite{Supplemental}. The resolution lower bound is $\Delta \delta_r\ge 1/\sqrt{\mathcal{I}(\delta_r)}$, where $\mathcal{I}(\delta_r)=(\partial_{\delta_r}P_t)^2/\sigma^2_{\mathrm{noise}}$ is the Fisher information (FI) \cite{gefen2019overcoming}. The variance of $P_t$ due to noise is $\sigma^2_{\mathrm{noise}}=\sigma^2_{\mathrm{QPN}}+\sigma^2_{\mathrm{readout}}$ with the main contributions typically coming from quantum projection noise (QPN) and photon shot noise affecting the readout, respectively \cite{Degen2017Quantum}. Then,  
\begin{equation}
\label{Eq:Precision-Main}
    (\Delta \delta_r)^2=\frac{\sigma^2_{\mathrm{QPN}}+\sigma^2_{\mathrm{readout}}}{(\partial_{\delta_r}P_t)^2}=\frac{\sigma^2_{\mathrm{QPN}}+\sigma^2_{\mathrm{readout}}}{4 b_{t}^2 \delta_r^2}.
\end{equation}
The effect of readout noise can be reduced significantly, e.g., by using single-shot-readout \cite{Neumann2010Single}, making QPN dominant \cite{Supplemental}. 
We carefully choose the interrogation times $t =2 n\pi /\omega_s $ (with $n$ a positive integer) to satisfy the superresolution condition \cite{gefen2019overcoming}. Then, $P_{t}\approx \Omega^2 t^2 \delta _r^2/\omega _s^2$ which results in $\sigma^2_{\mathrm{QPN}}=P_t(1-P_t)\approx P_{t}$ with the approximation valid for small $\delta_r\to 0$ and $t =2 n\pi /\omega_s $. 
Although both $P_t$ and $\sigma^2_{\mathrm{QPN}}$ approach zero as $\delta_r\to 0$, the FI does not vanish, taking the form 
\begin{equation}\label{eq:QFI_superresolution}
\mathcal{I}(\delta_r)=\frac{(\partial_{\delta_r}P_t)^2}{P_t(1-P_t)}\approx \frac{4\Omega^2 t^2}{\omega _s^2},
\end{equation}
so $\Delta\delta_r\ge \omega_s/(2\Omega t)$, which is finite for arbitrarily small $\delta_r$. This is also confirmed in  Fig.\,\ref{Principles_super} (b), which demonstrates that the local maximal difference between $P_t(\delta_{r}=0)$ and $P_t(\delta_{r}=0.05\omega_{s})$ indeed occurs when the superresolution conditions holds, namely $t = 2n\pi/\omega_{s}$. 
We note that the method is also applicable with conventional and other readout schemes \cite{Jelezko2004prl,Shields2015electric,Siyushev2019science,Barry2020rmp} as long as the readout noise is suppressed sufficiently, e.g., by repeating the experiment.

Despite its nonvanishing property, $\mathcal{I}(\delta_r)$ in Eq. \eqref{eq:QFI_superresolution} is not optimal, as the signal angular frequency $\omega_s$ can be quite large. It proves useful to apply a dynamical decoupling sequence of fast, ideally instantaneous, $\pi$ pulses with a pulse separation $\tau=\pi/\omega_{\text{DD}}$ in the sensing protocol \cite{Degen2017Quantum,gefen2019overcoming}. Dynamical decoupling improves the coherence time by applying an effective spectral noise filter with a characteristic sensing frequency $f_{\text{DD}}=1/(2\tau)$ \cite{Degen2017Quantum}. The time evolution of the system in the interaction basis of the pulses is given by the Hamiltonian 
\begin{equation}
H_{\text{eff}}\approx\sum_{i=1,2}\widetilde{\Omega}_i \sin[\delta_{i}t+\varphi_i (t)]\sigma_z,
\label{Eq:Hs-Effective}
\end{equation}
where $\delta_i=\omega_{\text{DD}}-\omega_i$ ($i=1,2$) with $\omega_{\text{DD}}=(2\pi)f_{\text{DD}}$, and $\widetilde{\Omega}=(2/\pi)\Omega$ is attenuated due to the pulses (see Fig. \ref{setup}(a) and \cite{Supplemental}). Thus, $\delta_s=\omega_{\text{DD}}-\omega_s$ and $\delta_r=(\delta_2-\delta_1)/2$ \cite{gefen2019overcoming}, as shown in Fig.\,\ref{setup}(a). The pulse separation is nearly resonant with the time-oscillating signal, i.e., $\delta_i\ll\omega_i$, resulting in $\mathcal{I}(\delta_r)\approx 4\widetilde{\Omega}^2 t^2/\delta_s^2$ and $\Delta\delta_r\ge \delta_s/(2\widetilde{\Omega}t)$, leading to improved resolution. As the superresolution condition requires $\delta_s =2 n\pi/t$, this leads to $\Delta\delta_r\ge (n\pi/\widetilde{\Omega})t^{-2}$, which is much better than the standard $t^{-1}$ scaling.

\emph{Experimental implementation.---}
We perform a proof-of-principle experimental implementation of the superresolution protocol using a single nitrogen-vacancy center (NV)  in diamond as a quantum sensor. The NV is positioned more than $20\,\mathrm{nm}$ below the diamond surface. The diamond has a hemispherical shape and is produced by chemical vapor deposition as a solid immersion lens \cite{Supplemental}, which enhances the excitation and detection efficiency 
\cite{siyushev2010monolithic,hadden2010strongly}. Specifically, we achieve a photon count of up to 1300 kcounts/s from this single NV center.
We use two independent radio frequency (RF) generators to generate two time-oscillating signals, which are sent to the diamond and interact with the NV electron spin. Its spin sublevels $m_{s}= -1$ and $m_{s}= 0$ shown in Fig.\,\ref{setup}(b) are selected to form an effective qubit sensor. 

After the interaction with the signal, we measure the NV sensor spin to extract the transition probability in Eq.\,\eqref{Eq:Population} and estimate the signal frequencies. Conventional optical readout is performed by detecting its electron spin-dependent fluorescence, but it is affected by photon shot noise. 
To overcome this limit, we read out the final state of the NV sensor by employing single shot readout (SSR)  \cite{jiang2009repetitive,Neumann2010Single}, as illustrated in Fig.\,\ref{setup}(c). We use the intrinsic $\prescript{15}{}{\mathbf{N}}$ nuclear spin as a memory qubit. In each measurement, the information about the electron spin state is first transferred onto the $\prescript{15}{}{\mathbf{N}}$ nuclear spin via an effective controlled-NOT (C$_{\text{e}}$NOT$_{\text{n}}$) gate with an electron spin selective $\pi_{\text{RF}}$ pulse on the nuclear spin. Then, the nuclear spin state is detected via SSR with a repetitive readout $R$ times.

%
\begin{figure}[t]
\centering
\includegraphics[width=8cm]{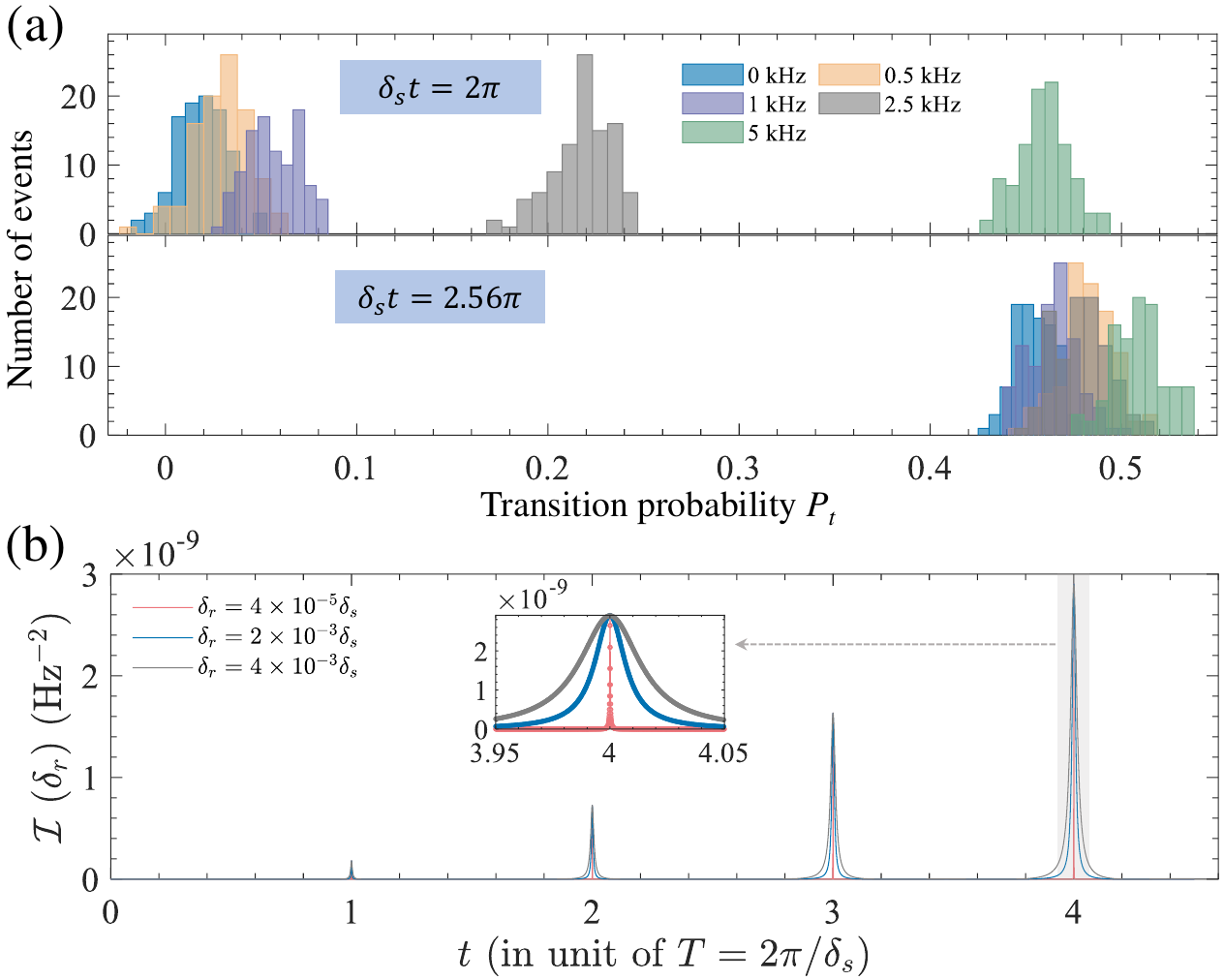}
\caption{ Demonstration of superresolution condition. (a) The histograms of experimentally measured $P_t$ for frequency separation: $\delta_{r}/2\pi=[0, 0.5, 1, 2.5, 5]\,$kHz. They show apparent (negligible) distinguishability when the superresolution condition is (not) satisfied. (b) Simulated Fisher information, which determines the estimation uncertainty of $\delta_r$, shows gradually increased peaks when the interrogation time $t$ approaches $2n\pi/\delta_s$. Here we set $\delta_{s}=(2\pi) 50\,$kHz, $\widetilde{\Omega}=(2\pi)16.85$ kHz.}
\label{results_optimal_condition}
\end{figure}
%

\begin{figure*}[t]
\centering
\includegraphics[width=0.8\linewidth]{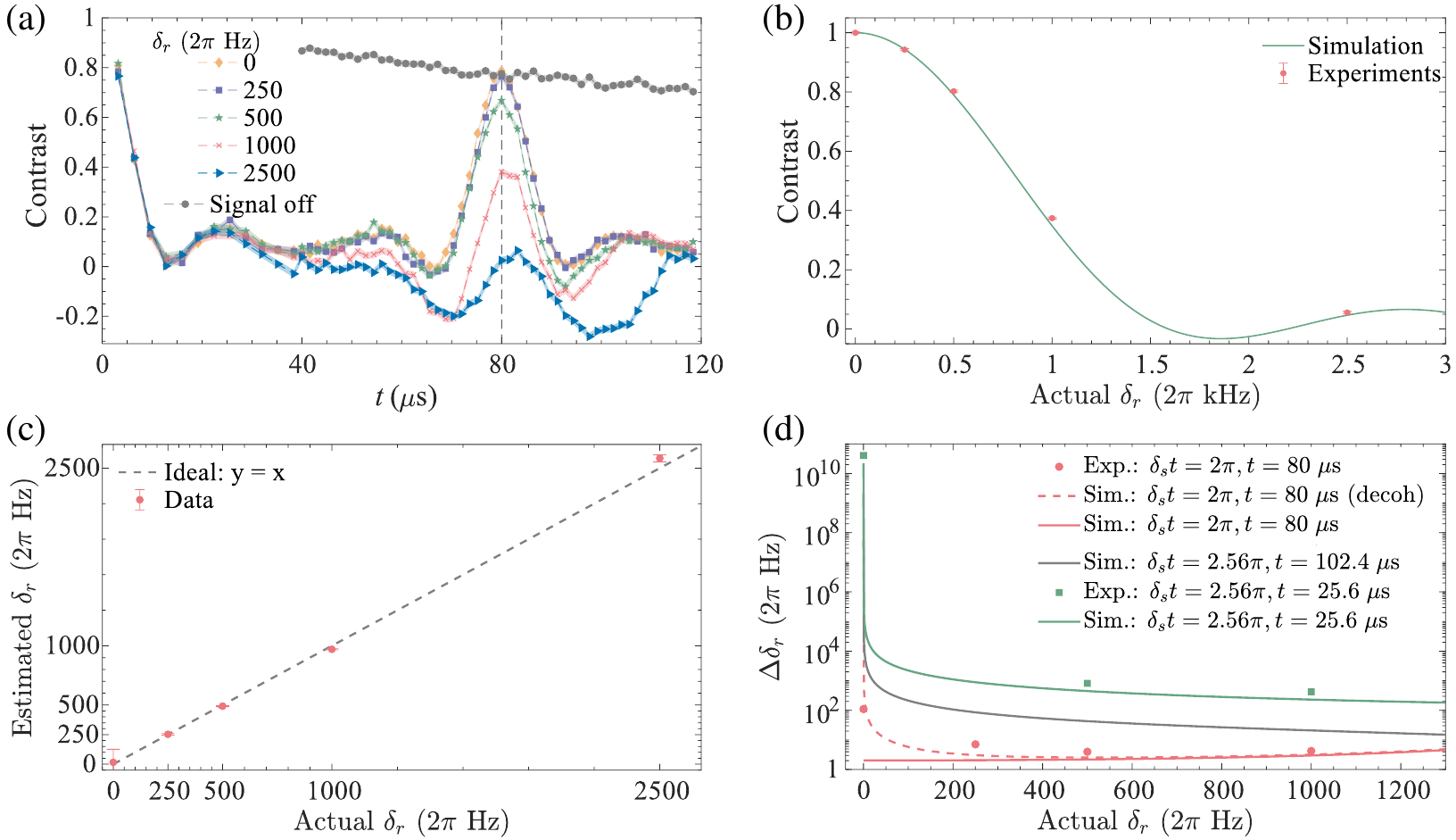}
\caption{(a) Signal contrast vs. total measurement time $t$ with dynamical decoupling with pulse separation $\tau=\pi/\omega_{\text{DD}}=200\,$ns for different $\delta_{r}$. 
The maximum difference is at the superresolution time of $t=2\pi/\delta_s=80\,\mu$s. (b) Experimental (red dots) and simulated (green line) signal contrast at $t=80\,\mu$s  vs. angular frequency difference $\delta_r$. 
(c) Estimated vs. actual frequency separation $\delta_{r}$. We note that the approximation $\delta_r\ll \delta_s$ is not valid for $\delta_r=(2\pi)\,2.5$ kHz. 
(d) Estimation uncertainty $\Delta\delta_r$ vs. $\delta_r$, which is consistent with Eq.\,\eqref{Eq:Precision-Main}. The green squares and line represent the measured and simulated $\Delta\delta_r$ at $t = 2.56\pi/\delta_s$ under the assumption of perfect measurements, corresponding to the data in Fig.\,\ref{results_optimal_condition}(a). The thin gray line shows simulations for an increased interaction time $t = 2.56\pi/\delta_s=102.4\,\mu$s. The red dots represent the experimental data at the superresolution condition. The red solid line shows the simulated $\Delta\delta_r$ derived from the theoretical Fisher information under the assumption of perfect measurements with the same experimental parameters. The red dashed line represents the simulated $\Delta\delta_r$ obtained by incorporating the decoherence of the quantum sensor \cite{Supplemental}, showing excellent agreement with the experimental results. 
}
\label{results_fit}
\end{figure*}

Experimentally, we apply a strong magnetic bias field of 598 mT generated by the superconducting magnet, and the $\prescript{15}{}{\mathbf{N}}$ nuclear spin average lifetime is extended to approximately 60 ms during single-shot readout, which includes laser illumination \cite{Supplemental}. Such a long lifetime allows us to achieve a large number of repetitive readouts of the nuclear spin state with a quantum non-demolition measurement. In our experiments, the nuclear spin readout to detect a given state in a single shot reaches a high fidelity of $\mathscr{F}=99.69\%$, as shown in Fig.\,\ref{setup}(d). We repeat the single measurement $M\thickapprox 4400$ times with a total measurement time of $20$ seconds and extract the averaged transition probability of the NV sensor in Eq.\,\eqref{Eq:Population} by analyzing the information on the flip rate of the nuclear spin. Our SSR-assisted method yields a nearly twofold enhancement in signal-to-noise ratio (SNR) compared to traditional photoluminescence readout \cite{Supplemental}, significantly suppressing classical readout noise (see Eq.\,\eqref{Eq:Precision-Main}).

\emph{ Demonstration of superresolution protocol.---}
In a first experiment to demonstrate the superresolution protocol, we generate two independent oscillating signals centered at $\omega_{s}=(2\pi)5.05$ MHz from the two RF generators with their amplitudes calibrated to be practically identical. The difference between their angular frequenices $\delta_{r}$ varies from $(2\pi)$0 to $(2\pi)$5 kHz. An XY8-$N$ pulse sequence, i.e., XY8, repeated $N$ times with a total of $8N$  $\pi$ pulses, is employed. The spacing of $\tau=100$ ns between the centers of the $\pi$ pulses results in a spectral filter centered at $\omega_{\text{DD}}=(2\pi)5$ MHz and subsequently an effective central frequency at $|\delta_s|=\omega_{s}-\omega_{\text{DD}}=(2\pi)50$ kHz. Thus, the superresolution condition can be fulfilled with an interrogation time of $t=2n\pi/\delta_s=20n\,\mu$s, corresponding to $N=25n$ repetitions of XY8. 

We compare the measurement of frequency separation at two different times of $P_t$, i.e. $ t=2\pi/\delta_s=20\,\mu$s and $ t=2.56\pi/\delta_s=25.6\,\mu$s. The estimates of $P_t$ are given by the histograms in Fig.\,\ref{results_optimal_condition} (a). When the measurement satisfies the superresolution condition $\delta_s t=2\pi$, it yields a significantly enhanced response to the frequency separation, allowing for distinguishability of the two frequencies. This is  
confirmed by the dynamics of Fisher information under the assumption of no classical measurement noise, namely $\mathcal{I}(\delta_r)=(\partial_{\delta_r} P_t)^2/\sigma^2_{\mathrm{QPN}}$, displaying prominent peaks at these positions, as shown in Fig.\,\ref{results_optimal_condition} (b).

\emph{Resolution of nearly identical frequencies.---}
In a second experiment, we estimate the sub-kHz frequency difference between two nearly identical incoherent signals. The center frequency of the two signals is now $\omega_{s}=(2\pi) 2.5125$ MHz and the amplitude is estimated  $\widetilde{\Omega}\approx(2\pi) 16.85$ kHz \cite{Supplemental}. The frequency separation takes the values $\delta_{r}=(2\pi)[0.0, 0.25, 0.5, 1.0, 2.5]$ kHz. Similarly, we apply the XY8-$N$ sequence with $\pi$ pulse spacing of $\tau=200\;$ns ($\omega_{\text{DD}}=(2\pi)2.5\,$ MHz), shifting the central frequency to $\delta_{s}=(2\pi)$12.5 kHz. We note that the minimum angular frequency detuning $\delta_s>\delta_{s,\text{min}}\sim (2\pi)/T_{2}$ is fundamentally constrained by the coherence time of the quantum sensor, where $T_{2}\approx 1.3$ ms for the particular XY8 dynamical decoupling sequence we use. 
In order to reduce errors due to charge-state and count-rate fluctuations, we alternate the phase of the final $\pi/2$ pulse to be $0^{\circ}$ or $180^{\circ}$ to obtain $1-P_{t}$ and $P_{t}$, respectively, and subtract them to obtain the signal contrast $C_t= (1-P_{t}) -P_{t}=\prod_{i=1}^2 J_{0} \left(\frac{4\widetilde{\Omega}}{\delta_i}\sin\left(\frac{\delta_i t}{2}\right)\right)$, where $\delta_i=\omega_{\text{DD}}-\omega_i$. In Fig.\,\ref{results_fit} (a), we show the evolution of $C_t$ vs. total interaction time. The contrast shows a maximal response to the frequency separation at $t=80\;\mu$s, 
which satisfies the superresolution condition $t= 2\pi/\delta_s$. 

Next, we fix the evolution time at $t=80\;\mu$s and perform $30$ measurements of $C_t$ for every $\delta_r$ to obtain an average, reducing the statistical uncertainty, as shown in Fig.\,\ref{results_fit} (b). The contrast is normalized to the case without a signal. The experimentally measured contrast $C_t$ fits excellently the theoretical values. 
Next, we estimate the frequency separation from these measured $C_t$, resulting in 
$\delta_r \approx \delta_s \sqrt{1-C_t}/(\sqrt{2}\widetilde{\Omega} t) $. 
Fig.\,\ref{results_fit} (c) demonstrates that the estimated frequency differences match excellently with their actual values. We note that we estimate the frequency separation using Eq. \eqref{Eq:Population} for large values of $\delta_r$, where the contrast is lower than $0.5$ \cite{Supplemental}. 
The uncertainty of the estimated frequency difference is 
\begin{equation}
\label{Eq:separation_error}
\Delta\delta_r \approx \frac{\delta _s}{2\sqrt{2}\widetilde{\Omega } t}\frac{\Delta C_t }{\sqrt{1-C_t}}=\frac{\pi}{\sqrt{2}\widetilde{\Omega} t^2}\frac{\Delta C_t}{\sqrt{1-C_t}},
\end{equation}
where we neglect the uncertainties in $\delta_s$ and $\widetilde{\Omega}$, which is feasible for small $\delta_r$ \cite{Supplemental}. We depict the estimation uncertainty (simulated  $\Delta\delta_r$ bounded by the FI) 
in Fig.\,\ref{results_fit}(d). 
The experimental data exhibit excellent agreement with the simulation. We note that FI does not remain constant for very small values of $\delta_r$ due to other noise sources apart from QPN, e.g., decoherence and residual classical readout noise, leading to an increase in the experimental estimation error. As shown in Fig. \,\ref{results_fit}(d), a simulation accounting for decoherence fits excellently the experimental values in this regime \cite{Supplemental}. 
Based on the latter simulation and the resolution criterion $\Delta\delta_r \le|\delta_r|$, we estimate the best achievable frequency resolution of ($2\pi$)23.3 Hz. The experimental data and simulations at $t=2.56\pi/\delta_s$ when the superresolution condition is not satisfied demonstrate a much higher estimation uncertainty, as expected from theory. The experiments clearly demonstrate the advantage of our protocol, allowing for significantly improved resolution, e.g., in nanoscale NMR. 

\emph{ Conclusion.---}
We demonstrate the principles of superresolution quantum sensing with a solid-state single spin quantum sensor by experimentally resolving two incoherent signals with nearly identical frequencies. We use single-shot readout, assisted by the intrinsic $\prescript{15}{}{\mathbf{N}}$ nuclear spin as memory to limit classical readout noise. The protocol nearly eliminates quantum projection noise by quantum control, leading to a finite Fisher information even when the frequencies are arbitrarily close.  
These findings establish a practical framework for high-resolution nano-NMR, advancing the capabilities of quantum spectroscopy in detecting closely spaced spectral features, which can be related to $J$ coupling or chemical shift.  
The uncertainty of the frequency separation improves as $\Delta \delta_r\sim t^{-2}$ with the superresolution method in comparison to $t^{-1}$ with standard approaches. Resolution can be improved even further with a prolonged interrogation time, which can reach more than a millisecond for electronic spin sensors \cite{Balasubramanian2009Ultralong,Herbschleb2019Ultra,Salhov2024Protecting}. We note that spin-squeezed or entangled states with ensembles of NV centers as quantum probes have already been demonstrated \cite{Cappellaro2009Quantum,Bennett2013Phonon,wu2025spinsqueezing} and 
could in principle be combined with our superresolution protocol for further improvement of sensitivity. 
Employing squeezed states of light has also allowed to surpass the standard quantum limit in other systems \cite{Jia2024Squeezing,Feng2025Quantum,Troullinou2021Squeezed} but these methods 
go beyond the scope of our work. 
We implement the method in the context of pulsed dynamical decoupling and Ramsey sequences, enabling detection frequencies from dc to several tens of MHz \cite{Degen2017Quantum}. Extending these techniques to higher-frequency domains in the GHz range might also be possible by using appropriate sequences \cite{CaiNJP2012,Salhov2024Protecting}, highlighting the broad applications of the method.

\emph{ Acknowledgment.---}
The authors thank Tuvia Gefen for useful discussions. This work was supported by the BMBF via projects QSENS, MikQSens and Diaqnos, EUREKA via project quNV2.0, QUANTERA via project “Microfluidics Quantum Diamond Sensor”, Carl Zeiss Foundation via IQST and Ultrasens-Vir projects, DFG via EXC 2154: POLiS, SFB 1279 and projects 491245864, 499424854, 387073854, ERC grant HyperQ (856432), EU via H2020 projects FLORIN SPINUS and QuMicro. J.-M. C. and Y.-M. C. are supported by the National Natural Science Foundation of China (12425414, 12304572) and Quantum Science and Technology-National Science and Technology Major Project (2024ZD0300900, 2024ZD0300902). Y. L. is supported by the BMBF under the funding program ‘quantum technologies—from basic research to market’ in the project Spinning (13N16215) and project CoGeQ (13N16101).  A.R. acknowledges the support of ERC grant QRES, project number 770929, Quantera grant MfQDS, the Israeli Science foundation (ISF), the Schwartzmann university chair and the Israeli Innovation Authority under the project ``Quantum Computing Infrastructures''.

%


\clearpage
\onecolumngrid
\setcounter{secnumdepth}{2} 
\makeatletter
\def\thesection{\Roman{section}}
\def\@hangfrom@section#1#2#3{#1#2#3}
\renewcommand{\@seccntformat}[1]{\csname the#1\endcsname.\quad}
\makeatother

\renewcommand{\thefigure}{S.\arabic{figure}}
\renewcommand{\theequation}{S.\arabic{equation}}
\setcounter{figure}{0}
\setcounter{equation}{0}
\setcounter{section}{0}

~\\
\begin{center}
	\textbf{\large -Supplemental Material- \\Overcoming frequency resolution limits using a solid-state spin quantum sensor}
\end{center}

~\\

\section{Superresolution Quantum Sensing}
In the context of quantum sensing, the information on the frequency parameters are encoded into a quantum state, which we denote by a density matrix $\rho$, through evolution governed by the Hamiltonian for a certain time. Stated by the so-called quantum  Cram\'er-Rao bound, the ultimate measurement precision of $\delta_r$ is constrained by 
\begin{equation}
\Delta \delta_r\geq I(\delta_r)^{-1/2}, 
\end{equation}
 where $I(\delta_r)$ is the associated quantum Fisher information (QFI). For a general quantum state spectrally decomposed as $\rho=\sum_{n} p_n |n\rangle\langle n|$, the QFI is defined by the following structure,
\begin{equation}
\label{Eq:QFI-Main}
    \mathcal{I}(\delta_r)=2\sum_{n,m}\frac{|\langle n|\partial_{\delta_r} \rho|m\rangle|^2}{p_n+p_m}.
\end{equation}
A key problem in frequency resolution is that the response of $\rho$ with respect to $\delta_r$ (i.e. $\partial_{\delta_r}\rho$) tends to vanish if the two signals are close enough, i.e. $\delta_r\rightarrow 0$ with nearly identical frequencies and amplitudes \cite{rotem2019limits}. 
Therefore, it is likely that $I(\delta_r)\to 0$ which subsequently indicates a vanishing resolving ability for two nearly identical time-oscillating signals.

It is remarkable that a nontrivial case will emerge where a certain eigenvalue of $\rho$ in the denominator of Eq.\,\eqref{Eq:QFI-Main} goes to zero as well when $\delta_r\rightarrow 0$, which in principle can be achieved by controlling the dynamics of quantum sensing process. This makes it possible to eliminate the singularity in the quantum  Cram\'er-Rao bound \cite{tsang2016quantum,nair2016far,lupo2016ultimate,chrostowski2017super,Chu2023PRL,Chu2024PRL}, and in turn to overcome the resolution limit. As an example, for $p_i\sim \delta_r^\alpha$, we have 
\begin{equation}\label{Eq:QFI-bound}
\mathcal{I}(\delta_r) \geq (\partial_{\delta_r} p_i)^2/p_i\sim \delta_r^{\alpha-2} ,
\end{equation}
 which gives a non-vanishing value if $1<\alpha\leq 2$, even for the case where $\partial_{\delta_r} \rho = 0$ \cite{gefen2019overcoming}.

~\\

\section{Derivation of Transition Probabilities}

For signals given by Eq.(1) in the main text, for example, the two signals generated by two independent generators in our experiments, the amplitudes are constant while the phases of the two completely incoherent signals follow two independent uniform distributions, i.e., $\Omega_{1,2}=\Omega$, $\varphi_i\sim U[0,2\pi]$. Significantly enhanced signal detection can be achieved by applying a suitable sequence of $\pi$ pulse that periodically flip the NV spin sensor and thus effectively change the frequency of oscillations. For simplicity, we first consider the Hamiltonian of one signal as follows 
\begin{equation}
H=\Omega\sin(\omega t+\varphi)\sigma_z. 
\end{equation}
The periodic $\pi$ pulses are applied at a frequency of $\omega_{\text{DD}}=\omega+\delta$ (namely the duration between $\pi$ pulses is $\pi/\omega_{\text{DD}}$, and $\delta$ is referred to as detuning) on the NV spin sensor. This results in the following effective Hamiltonian \cite{gefen2019overcoming} as
\begin{equation}\label{Heff_1}
   H_{\text{eff}}=\tan\left(\frac{\pi}{2(1+\delta/\omega)}\right)\left(\frac{\delta}{\omega}\right)\Omega\sin(\delta t+\varphi)\sigma_z.
\end{equation}
As can be seen, the control pulses effectively shift the signal's frequency from $\omega$ to $\delta$, and scale the amplitude by a factor of $\tan\left(\frac{\pi}{2(1+\delta/\omega)}\right)\left(\frac{\delta}{\omega}\right)$. Usually, the regime $\delta\ll\omega$ is satisfied, and thereby $\tan\left(\frac{\pi}{2(1+\delta/\omega)}\right)\left(\frac{\delta}{\omega}\right)\approx\frac{2}{\pi}$, which subsequently leads to that
\begin{equation}\label{Heff_2}
	H_{\text{eff}}\approx \frac{2}{\pi}\Omega\sin(\delta t+\varphi)\sigma_z.
\end{equation}
\begin{figure}[hpbt]
\centering
\includegraphics[width=0.6\linewidth]{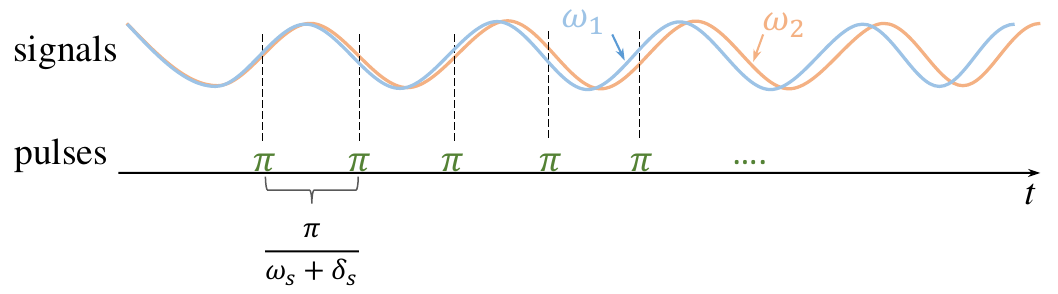}
\caption{ {\bf Periodic control pulses to enhance signal detection.} We assume the central frequency of two signals $\omega_{1,2}$ is $\omega_{s}$, and the periodic $\pi$ pulses are applied at a frequency of $\omega_{\text{DD}}=\omega_s+\delta_s$.}
\label{pulse_scheme}
\end{figure}
One can find that the above derivation can be straightforwardly generalized to the scenario of two incoherent signals with frequencies ($\omega_1,\omega_2$), and the effective Hamiltonian reads, 
\begin{equation}\label{Heff_two}
	H_{\text{eff}}\approx \sum_{i}\frac{2}{\pi}[\Omega_i\sin(\delta_i t+\varphi_i)]\sigma_z.
\end{equation}
where $\delta_i=\omega_{\text{DD}}-\omega_{i}$ with $i=1,2$. By further introducing the central frequency of the two signals as $\omega_s=(\omega_1+\omega_2)/2$, as shown by Fig.\,\ref{pulse_scheme}, one obtains
\begin{equation}\label{delta_r}
	\begin{aligned}
	&\delta_s=\omega_{\text{DD}}-\omega_s=\frac{\delta_1+\delta_2}{2}, \\
	&\delta_r=\frac{\delta_2-\delta_1}{2}=\frac{\omega_1-\omega_2}{2}.
\end{aligned}
\end{equation}

\begin{figure}[b]
\centering
\includegraphics[width=0.85\linewidth]{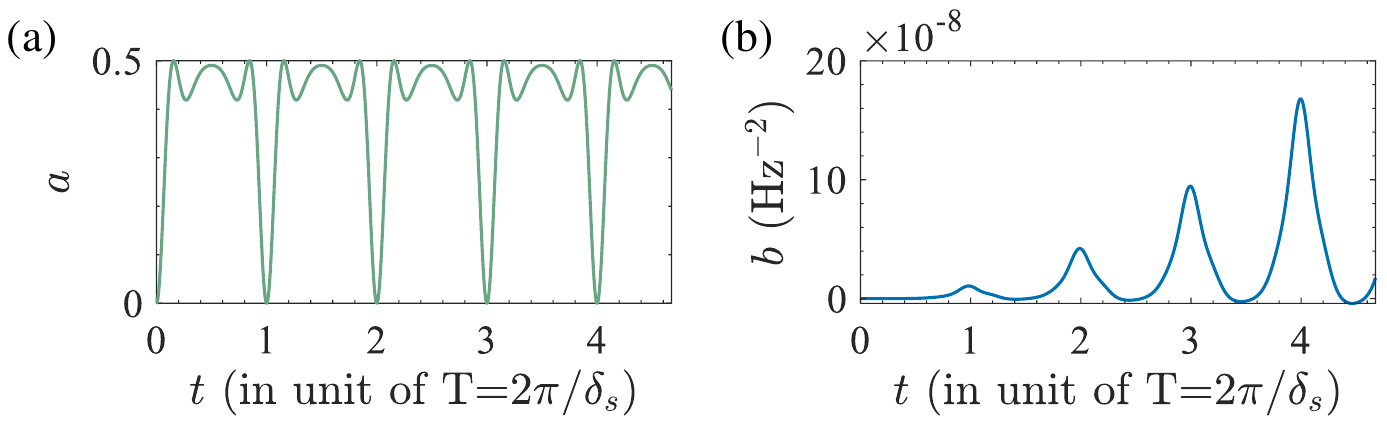}
\caption{{\bf Coefficients of second-order expansion of the transition probability.} (a) The zeroth-order coefficient $a_t$ of the transition probability versus $\delta_r$. (b) The second-order coefficient $b_t$ of the transition probability versus $\delta_r$. Here $\delta_s=(2\pi)12.5$ kHz was selected for simulation.
}
\label{coefficients}
\end{figure}

Below we consider the case where $\Omega_1=\Omega_2=\Omega$ and set $\widetilde{\Omega}=\frac{2}{\pi}\Omega$, the Hamiltonian in Eq.\,\eqref{Heff_two} can then be rewritten as 
\begin{equation}\label{Heff_two_phase}
	H_{\text{eff}}\approx \sum_{i}\widetilde{\Omega}\sin(\delta_i t+\varphi_i)\sigma_z.
\end{equation}
In standard Ramsey interferometry by preparing the sensor along the positive $x$ axis and letting it evolve under the Hamiltonian in Eq.\,\eqref{Heff_two_phase}, the time-evolved quantum state is given by $(e^{-i\phi}|0\rangle+e^{i\phi}|1\rangle)/\sqrt{2}$, with the accumulated phase, $\phi$, which is given by
\begin{equation}\label{phase2_xy8}
	\begin{aligned}
		\phi
		\approx \sum_{i} \frac{\widetilde{\Omega}[\cos(\varphi_i)-\cos(\delta _i t+\varphi_i)]}{\delta_i}  .
	\end{aligned}
\end{equation}
The average transition probability over the random initial phases of the incoherent signals can then be derived as
\begin{align}\label{pa_phase}
	P_t&= \langle\sin^2(\phi)\rangle_{\varphi_i}= \left\langle\frac{1-\cos(2\phi)}{2}\right\rangle_{\varphi_i} \notag \\
	&=\frac{1}{2}-\frac{1}{2}\frac{1}{(2\pi)^2}\int_{0}^{2\pi}\int_{0}^{2\pi} \cos(2\phi)  d\varphi_1 d\varphi_2   \notag\\
	&=\frac{1}{2}\left[1-J_{0}\left(\frac{4\widetilde{\Omega}}{\delta_{1}}\sin\left(\frac{\delta_{1}t}{2}\right)\right) J_{0}\left(\frac{4\widetilde{\Omega}}{\delta_{2}}\sin\left(\frac{\delta_{2}t}{2}\right)\right) \right],
\end{align}
where $J_{0}(x)$ denotes the first-kind zeroth-order Bessel function. 
For small values of $\delta_r$, we can expand the transition probability to its second order as follows
\begin{equation}
P_t\approx a_t+b_t\,\delta_r^2, 
\end{equation}
with $a_t$ and $b_t$ functions of $\{\widetilde{\Omega},\delta_s,t\}$ as follows: 
\begin{equation}\label{a_formula}
	a_t = \frac{1}{2} \left[ 1 - J_0\left( \frac{4 \widetilde{\Omega} \sin\left(\frac{\delta_s t}{2}\right)}{\delta_s} \right)^2 \right],
\end{equation}

\begin{equation}\label{b_formula}
\begin{aligned}
b_t =&
\frac{\widetilde{\Omega}}{2 \delta_s^4}  \left\{
 4 \widetilde{\Omega} \left[ J_0^2 \left( \frac{4 \widetilde{\Omega} \sin\left(\frac{ \delta_st}{2}\right)}{\delta_s} \right) 
+ J_1^2 \left( \frac{4 \widetilde{\Omega} \sin\left(\frac{\delta_st}{2}\right)}{\delta_s} \right) \right]
\left[-2 +  \delta_s t \cot\left(\frac{\delta_s t}{2}\right) \right]^2 \right.\\
&\left. -  \delta_s J_0 \left( \frac{4 \widetilde{\Omega} \sin\left(\frac{ \delta_s t}{2}\right)}{\delta_s} \right) 
J_1 \left( \frac{4 \widetilde{\Omega} \sin\left(\frac{\delta_s t}{2}\right)}{\delta_s} \right) 
\left(-2 + \delta_s^2 t^2 + 2 \cos(\delta_s t) \right) \csc^3\left(\frac{\delta_s t}{2}\right)  \right\}
\sin^2\left(\frac{ \delta_s t}{2}\right),
\end{aligned}
\end{equation}
where $J_0(.)$ and $J_1(.)$ are the zeroth and first order Bessel function. The values of $a_t$ and $b_t$ changing with $t$ are shown in Fig.~\ref{coefficients}, which shows periodic dips
or peaks at the superresolution times $\tau =2 n\pi /\delta_s $ (with $n$ a positive integer).

\begin{figure}[hpbt]
\centering
\includegraphics[width=0.65\linewidth]{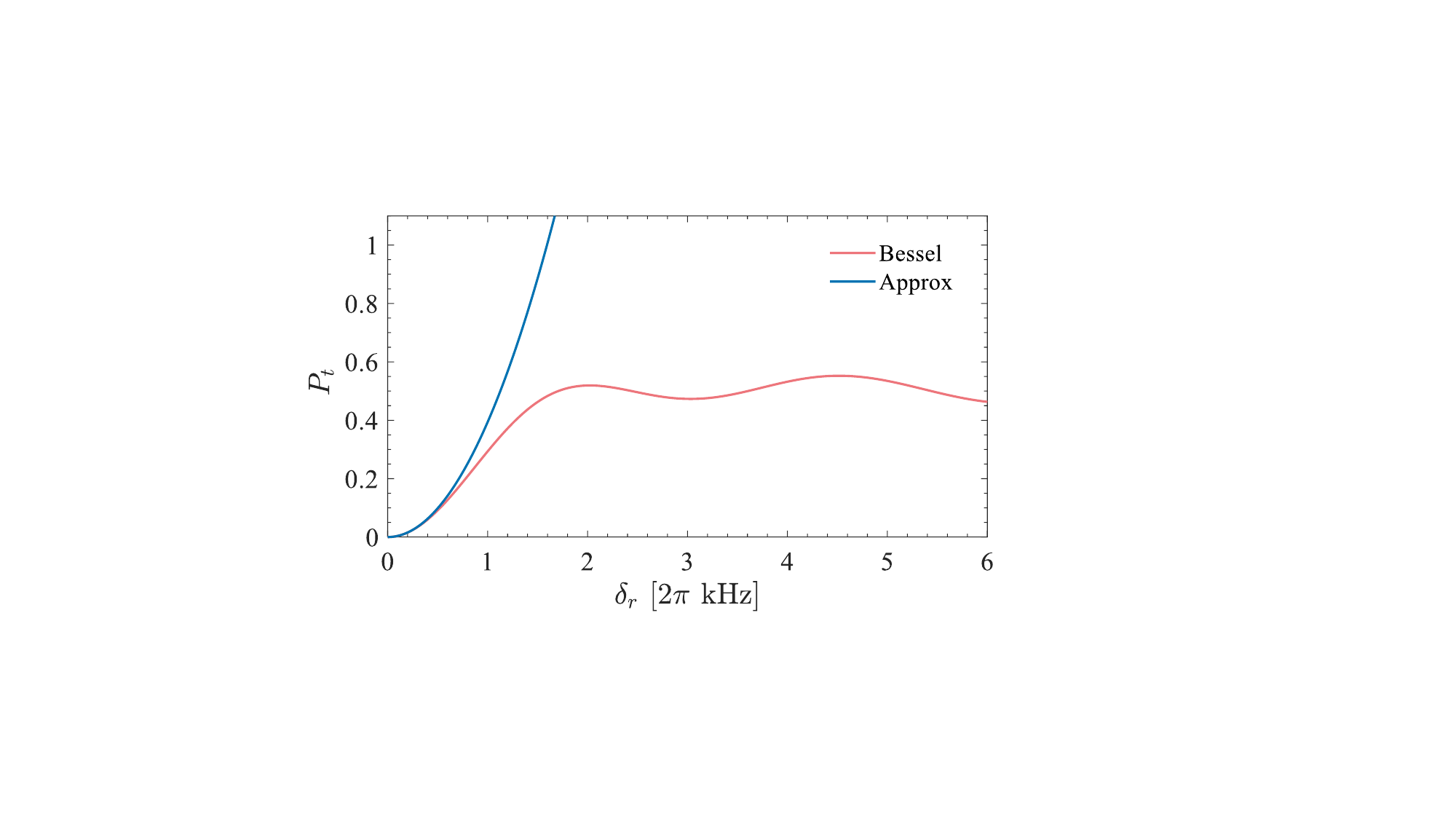}
\caption{{\bf Function curves of $P_t$
with respect to $\delta_r$.} Function curves of $P_t$
with respect to $\delta_r$ obtained by Eq. \eqref{pa_phase} (red curve) and Eq. \eqref{phase2_p_xy8} (blue curve). This indicates that the two functions are approximately equal only when $\delta_r < (2\pi)$1 kHz. This explains the deviation between the estimated and real $\delta_r$ when $\delta_r > (2\pi)$1 kHz in the Fig. 4(c) of the main text.
}
\label{P_difference}
\end{figure}

When the superresolution condition is satisfied, i.e. $\delta_s t=2\pi$ (see Fig.~\ref{coefficients}), we have $a_t=0$, $b_t=\widetilde{\Omega}^2t^4/(2\pi)^2$. Accordingly, the phase and average transition probability for $\delta_rt\ll1$ can be approximated as
\begin{equation}\label{phase2_p_xy8}
	\phi\approx\widetilde{\Omega}\delta_r t \frac{[\sin(\varphi_1)-\sin(\varphi_2)]}{\delta_s} \;\;\to\; P_t=b_t\delta_r^2\approx (\frac{\widetilde{\Omega}t}{2\pi})^2\delta_r^2 t^2.
\end{equation}
In the absence of the classical readout noise $\sigma_{\mathrm{readout}}$, the Fisher information is thus given by 
\begin{equation}\label{Ir_mixed}
 \mathcal{I} (\delta_r)=\frac{(\partial_{\delta_r} P_t)^2}{P_t(1-P_t)}=\frac{4\widetilde{\Omega}^2t^4}{(2\pi)^2}.
\end{equation}

Note that the approximations in Eq. \eqref{phase2_p_xy8} are only valid for $\delta_rt\ll1$. Using
our experimental parameters $\{\Omega_,\delta_s,t\}$, we plotted the function curves of $P_t$ with respect to $\delta_r$ obtained by Eq. \eqref{pa_phase} and Eq. \eqref{phase2_p_xy8} when the superresolution condition is satisfied, as shown in Fig.\,\ref{P_difference}. The two curves exhibit strong agreement within the range $\delta_r = (2\pi)$0 to $(2\pi)$1 kHz, but begin to diverge when $\delta_r > (2\pi)$1 kHz. This divergence explains the discrepancy observed between the estimated and the real $\delta_r$ when $\delta_r \geq (2\pi)$1 kHz in the Fig. 4(c) of the main text.
~\\


\section{Sample and Experimental Setup}

We performed the experiments on a single NV center spin in diamond. The sample is the electronic grade diamond with $[100]$ flat surface and natural $\prescript{13}{}{\mathbf{C}}$ isotopic content, which is polished into a hemispherical shape with a 2 mm diameter. The flat surface is overgrown with a 100 nm layer of isotopically enriched $\prescript{12}{}{\mathbf{C}}$ (99.999$\%$), using plasma enhanced CVD (chemical vapor deposition) \cite{osterkamp2015stabilizing}. The method of
delta-doping is used to create NV centers in the $\prescript{12}{}{\mathbf{C}}$ enriched layer. The depth of the single NV center
under investigation is estimated to be more than 20 nm. The diamond’s hemispherical shape acts as a solid immersion lens (SIL), enhancing the photon collection efficiency \cite{hadden2010strongly,siyushev2010monolithic}. For the NV center used in this work, we achieve a net photon flux up to $\sim1300$ kcounts/s.

\begin{figure}[hpbt]
\centering
\includegraphics[width=1.0\linewidth]{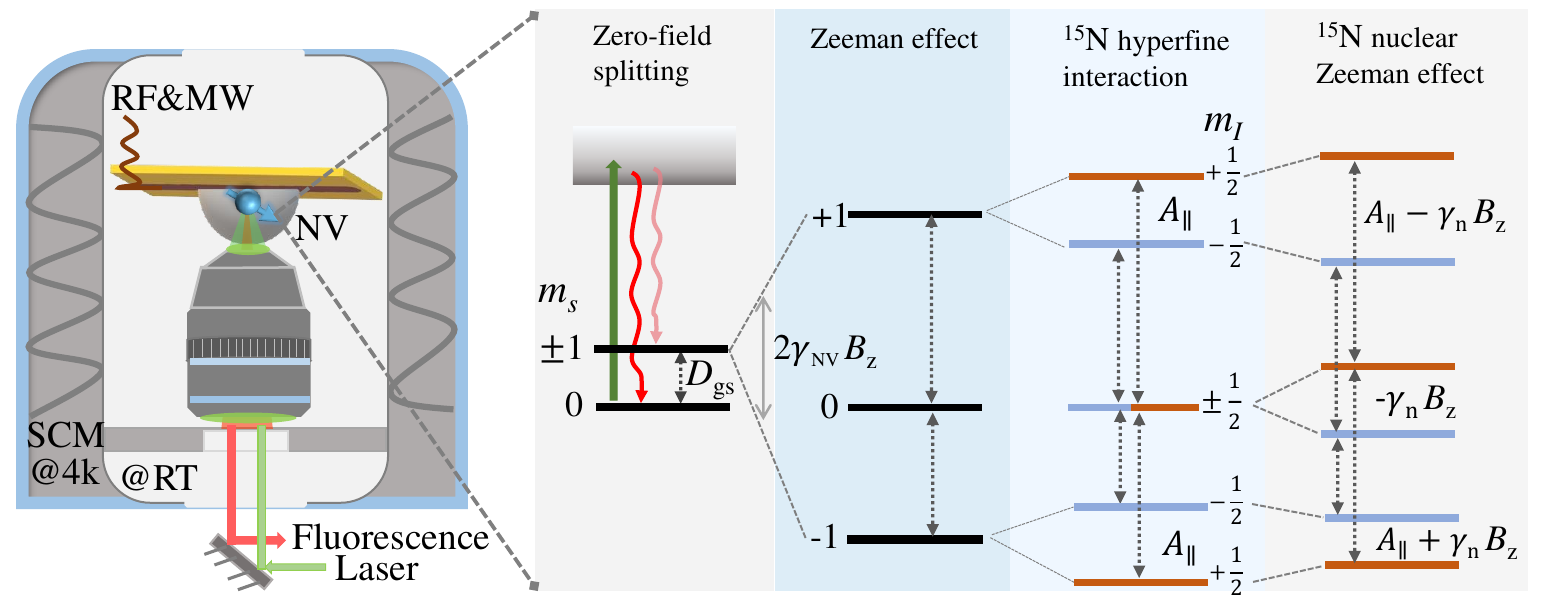}
\caption{ {\bf Schematic of the experimental setup and level scheme of the NV center’s electronic spin in the ground state.} The NV center in a hemispherical diamond SIL serves as a quantum sensor. The diamond is located in the room-temperature bore of the superconducting vector magnet (SCM). Due to the large applied external magnetic field $B_{z}$, nuclear Zeeman interactions are also taken into account in the energy level system. The relevant parameters include the zero-field splitting $D_{\text{gs}} =(2\pi) 2.87$ GHz, the gyromagnetic ratio of the NV electron spin $\gamma_{\text{NV}}=(2\pi)2.802$ MHz/G, the parallel hyperfine coupling strength with $\prescript{15}{}{\mathbf{N}}$ nucleus $A_{||}=(2\pi)3.03$ MHz, and the gyromagnetic ratio of $\prescript{15}{}{\mathbf{N}}$ nucleus $\gamma_{\text{n}}=(2\pi) -0.432$ kHz/G. Allowed transitions between energy levels are represented by dashed arrows.}
\label{setup_scheme}
\end{figure}

Experiments are performed on a home-built confocal setup. A schematic of the experimental setup is shown in Fig.\,\ref{setup_scheme}. A 532 nm laser (Laser Quantum gem 532) is used to initialize and readout the NV center spin states. An air objective mounted on a 3D piezo scanner realizes positioning single NV centers by focusing the 532 nm laser to the diamond sample and collecting the fluorescence response. The diamond sample, scanning components, and the objective are located in the room-temperature bore of a superconducting vector magnet (American Magnetics MX-311-40). With the capability of providing a tunable magnetic field up to 1 Tesla on the $x$, $y$ and $z$-axis, this magnet can create a magnetic field along arbitrary orientation at the bore center. Resonant microwave (MW) and radio-frequency (RF) pulses are generated by an arbitrary waveform generator (AWG, Tektronix AWG70001A, sampling rate 50 GSamples/s) with arbitrary phases and amplitudes.

Two independent signal generators (Rohde$\&$Schwarz SMB100A, Agilent Technologies 33522B) are used to generate two independent artificial signals, which are combined via an RF splitter. These signals are further combined with RF pulses from the AWG through an additional RF splitter, then amplified, and finally merged with MW pulses using a diplexer. The combined signals are transmitted to a coplanar waveguide antenna on a sapphire substrate, which is attached to the SIL sample. Photon detection is managed through a Time-Tagger single-photon counting card (Fast ComTec MCS6A), which receives photon events from avalanche photodiodes (APDs). All experimental procedures are controlled using the QuDi software suite \cite{BINDER201785}.

~\\
\section{Sensing and Readout Protocol}

\begin{figure}[hpbt]
\centering
\includegraphics[width=1.0\linewidth]{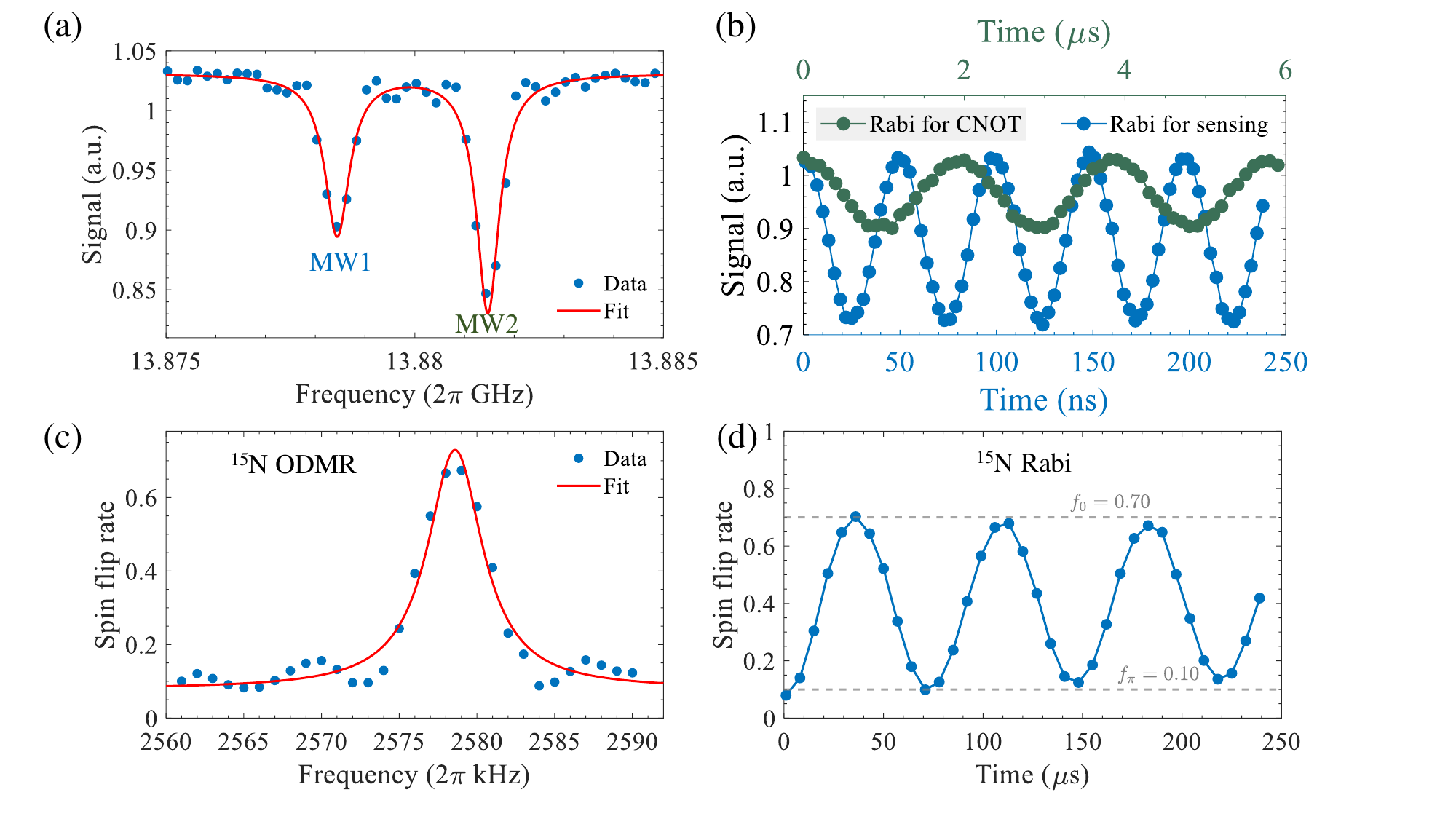}
\caption{ {\bf Realization of CNOT gates.} (a) Pulsed ODMR measurement of the $m_{S}= -1$ and $m_{S}=0$ transition showing the hyperfine splitting due to
the $\prescript{15}{}{\mathbf{N}}$ nucleus. The two probing frequencies MW1 and MW2 are indicated, which are used as C$_{n}$NOT$_{e}$ gate in SSR to drive electronic transitions selectively. (b) Electron Rabi measurements. In order to implement C$_{n}$NOT$_{e}$ gate, MW1 and MW2 are generated with a low MW power which shows a slow Rabi frequency (green data). In quantum sensing, spin manipulation is realized by high-power microwave pulses which drive electronic transitions regardless of the nuclear state (blue data). (c) ODMR measurement of $\prescript{15}{}{\mathbf{N}}$ nuclear spin for $m_{S} = 0$ spin state. This indicates that the RF excitation of a frequency $(2\pi)$ 2577 kHz can drive nuclear spin transitions ($|0\uparrow\rangle\leftrightarrow|0\downarrow\rangle$) as a C$_{e}$NOT$_{n}$ gate. (d) Rabi oscillation of $\prescript{15}{}{\mathbf{N}}$ nuclear spin with a frequency of $(2\pi)$ 13.8 kHz. The flip probability does not reach either 0 or 1 due to imperfect initialization of the NV center, restricted lifetime of the memory spin, and NV charge state conversion processes \cite{Schmitt2021Optimal}.}
\label{ODMR_Rabi}
\end{figure}

Figure~\ref{setup_scheme} shows the level scheme of the NV centre’s electronic spin in the ground state. The spin triplet states of the NV center are non-degenerate, with the $m_{S}= \pm1$ states shifted by the zero-field splitting $D_{\text{gs}} =(2\pi) 2.87$ GHz relative to the $m_{S}= 0$ state. With an external magnetic field $B_{z}=598$ mT applied along the NV axis, the degeneracy between the $m_{S}= \pm1$ energy levels is lifted by a Zeeman splitting of $2 \gamma_{\text{NV}} B_{z}$, where $\gamma_{\text{NV}}=(2\pi)2.802$ MHz/G is the gyromagnetic ratio of the NV electron spin. We selected the two states ($m_{S}= -1$, $m_{S}= 0$) as an effective qubit. Hyperfine coupling to the $\prescript{15}{}{\mathbf{N}}$ nuclear spin further splits each electronic state into sublevels, $m_{I}= -\frac{1}{2}$ and $m_{I}= +\frac{1}{2}$. Thus the two sublevels of $m_{S}= -1$ states exhibit a splitting of $A_{||}=(2\pi)3.03$ MHz, which represents the parallel hyperfine coupling strength. The sublevels are further lifted by nuclear Zeeman interactions due to the applied high magnetic field, resulting in a splitting of $\gamma_{\text{n}}B_{z}$, where $\gamma_{\text{n}}=(2\pi) -0.432$ kHz/G. The gray dashed lines with arrows in Fig.~\ref{setup_scheme} indicate allowed transitions between different $m_{S}$ states under resonant microwave driving. Consequently, two transition frequencies exist between the $m_{S}= -1$ and $m_{S}= 0$ states, as indicated by MW1 and MW2 in Fig.\,2(b) of the main text. A pulsed optically detected magnetic resonance (ODMR) measurement can resolve this hyperfine splitting, as illustrated in Fig.\,\ref{ODMR_Rabi}(a). With a low microwave amplitude, the $\pi$ pulses of MW1 and MW2 implement two C$_{n}$NOT$_{e}$ gates, which selectively flip the electron spin states depending on the nuclear spin states:
\begin{equation}
|0\downarrow\rangle\stackrel{\text{MW1}}\Longleftrightarrow |-1\downarrow\rangle, \;|0\uparrow\rangle\stackrel{\text{MW2}}\Longleftrightarrow |-1\uparrow\rangle . 
\end{equation}
The Rabi oscillations of electron spin, used to realize the C$_{n}$NOT$_{e}$ gates, are shown in Fig.\,\ref{ODMR_Rabi}(b) (green data), where the length of $\pi_{\text{MW1}}$ (or $\pi_{\text{MW2}}$) is about 1 $\mu$s. For quantum sensing, coherent control of the electron spin is achieved using pulses with a strong microwave amplitude, and the corresponding Rabi oscillations are depicted in Fig.~\ref{ODMR_Rabi}(b) (blue data). 
Since the C$_{e}$NOT$_{n}$ gate flips the nuclear spin only when the NV electron spin is in the $|0\rangle$ state after the sensing part, the nuclear spin flip probability directly correlates with the population of the $m_{S}= 0$ state. A higher nuclear spin flip probability represents a higher population in the $m_{S}= 0$ state.

\begin{figure}[hpbt]
\centering
\includegraphics[width=1.0\linewidth]{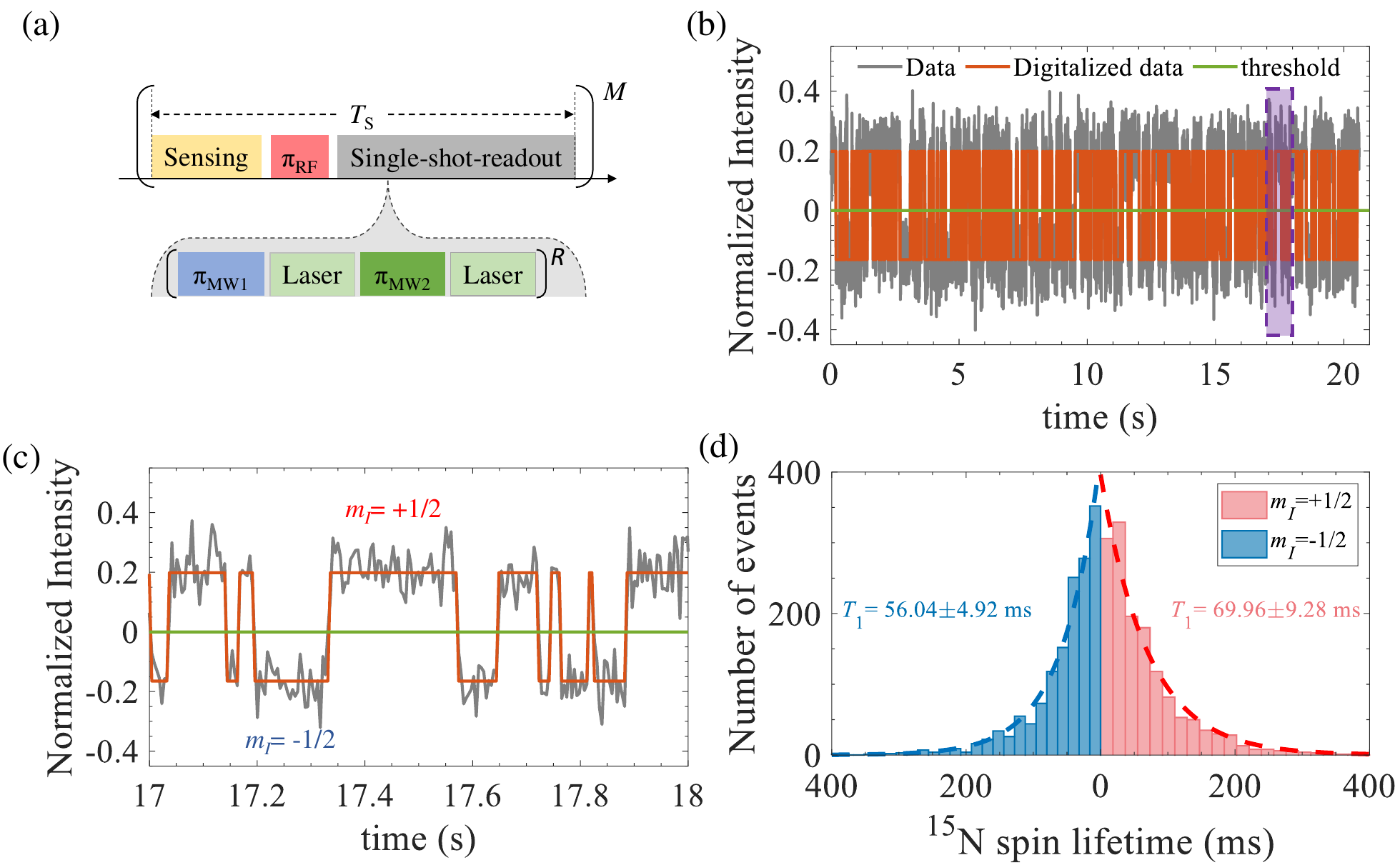}
\caption{ {\bf Single-shot readout of the single $\prescript{15}{}{\mathbf{N}}$ nuclear spin.} (a) SSR-assisted measurement scheme. One run of the whole procedure takes a running time $T_{S}$, and it's repeated for 20 seconds leading to a repetitions $M$. (b) Normalized fluorescence $I_{\text{norm}}$ time traces obtained by SSR measurement showing quantum jumps of a single nuclear spin in real time. The raw data is depicted in gray, while the digitalized data is represented by the orange line, obtained by applying a threshold indicated by the green line. A zoomed-in time trace of the dashed part is shown in (c), which provides clear evidence of these quantum jumps. (d) Statistic of the nuclear spin lifetime. The durations of nuclear spin-up and spin-down segments are separately counted and represented as histograms.}
\label{SSR_count_trace}
\end{figure}

The $^{15}{\mathbf{N}}$ nuclear spin lifetime is significantly improved under a high magnetic field \cite{Neumann2010Single}. Leveraging the extended nuclear spin lifetime and an improved photon count rate, we implement single-shot readout (SSR) to enhance the readout of the NV center's electronic spin state \cite{Neumann2010Single,haberle2017nuclear,Schmitt2021Optimal,Liu2017Single}. In our experiments, we choose the intrinsic $\prescript{15}{}{\mathbf{N}}$ nuclear spin as a memory spin. Fluorescence fluctuations can occur during repeated measurements due to unstable laser power or shallow NV centers, therefore, we implemented a normalized version of SSR \cite{haberle2017nuclear}. A schematic of this method is shown in Fig.\,\ref{SSR_count_trace}(a). First, the sensing experiment is conducted on the NV electron spin. Subsequently, the electron spin information is transferred onto the nuclear spin using a controlled NOT (C$_{e}$NOT$_{n}$) gate, implemented through a radio frequency (RF) $\pi$ pulse that flips the nuclear spin conditionally on the electron spin state. We characterized this C$_{e}$NOT$_{n}$ gate by performing the ODMR and Rabi measurements of $^{15}{\mathbf{N}}$ nuclear spin, as shown in Fig.\,\ref{ODMR_Rabi}(c,d), where the length of $\pi_{\text{RF}}$ is about 37 $\mu$s. Finally, the SSR is repeated $R$ times to readout the nuclear spin. The basic unit of normalized-SSR consists of two C$_{n}$NOT$_{e}$ gates (MW1 and MW2), and each is followed by a laser pulse, allowing probing of two nuclear spin states. These subsequent laser pulses can readout the NV electron spin and pump it into the $|0\rangle$ state. Due to the extended nuclear spin lifetime, while the NV center's electronic spin is repetitively pumped into the $m_{s}=0$ state, the memory ($^{15}{\mathbf{N}}$ nuclear spin) state is only weakly perturbed, such that the photon counts
from each readout window can be added. This enables the repetitive mapping of the nuclear spin state back onto the electron spin (repeated $R$ times), thereby allowing for high-fidelity measurement of the nuclear spin state in a single-shot readout. The accumulated photon counts after $R$ repetitions of the first and second laser pulses are recorded individually as $I_1$ and $I_2$, respectively, and a normalized count, $I_{\text{norm}}$, is calculated as follows\cite{haberle2017nuclear,Schmitt2021Optimal}:
\begin{equation}
I_{\text{norm}}=\frac{I_1 - I_2}{I_1 + I_2} .
\end{equation}
Using this approach, a low photon count is obtained if the $\prescript{15}{}{\mathbf{N}}$ nuclear spin state is $|{\downarrow}\rangle$, while a high photon count is obtained if the state is $|{\uparrow}\rangle$. The whole measurement sequence is repeated $M$ times, resulting in a time trace of $I_{\text{norm}}$.

The number of repetitions, $R$, is constrained by the backaction from the measurement that eventually flips the nuclear spin. In our experiment, we selected an optimal value of $R = 700$, and performed the full set of measurements with approximately $M \approx 4400$ repetitions. An obtained time trace of $I_{\text{norm}}$ is shown in Fig.\,\ref{SSR_count_trace}(b). A zoom in view of a portion of this trace is provided in Fig.\,\ref{SSR_count_trace}(c), which reveals quantum jumps of a single $^{15}{\mathbf{N}}$ nuclear spin in real time. The histogram of photon counts for $I_{\text{norm}}$ displays a well-distinguishable double-Gaussian distribution, as shown by Fig.\,2(d) of the main text. Due to the use of normalized SSR, the optimal threshold value is near 0, thus negative values of $I_{\text{norm}}$ correspond to the nuclear spin state $|{\downarrow}\rangle$, while positive values correspond to $|{\uparrow}\rangle$. A nuclear spin readout fidelity, $\mathscr{F}=99.69\%$ is estimated by fitting this double-Gaussian distribution. In the time trace of $I_{\text{norm}}$ shown in Fig.\,\ref{SSR_count_trace}(b), the durations of nuclear spin-up and spin-down segments are separately counted and represented as histograms in Fig.\,\ref{SSR_count_trace}(d). The distribution of these durations exhibits consistency with exponential decay, allowing for the derivation of lifetimes for the corresponding spin states through fitting procedures. The results indicate that the lifetime of the $\prescript{15}{}{\mathbf{N}}$ nuclear spin is approximately 60 ms at a magnetic field strength of 598 mT. Note that this is a mixture of the lifetime under laser illumination and the one without \cite{Neumann2010Single}.

~\\
\section{SNR Enhancement}

The signal-to-noise ratio (SNR) is commonly used to compare the level of a desired signal to the level of background noise. In a standard readout of the NV center, where only a laser pulse is used, the SNR can be approximated by the following expression \cite{haberle2017nuclear}:
\begin{equation}
\text{SNR}_{\text{std}}=\sqrt{\frac{n}{2}}\frac{c_{|0\rangle} - c_{|-1\rangle}}{\sqrt{c_{|0\rangle}^2 + c_{|-1\rangle}^2}}
\end{equation}
where $c_{|0\rangle}$ is the normalized signal corresponding to the state  $|0\rangle$, and $c_{|-1\rangle}$ is the normalized signal for the state $|-1\rangle$. The variable $n$ denotes the total number of photons accumulated during the detection window of the laser readout with $N_{\text{sweep}}$ repetitive measurements. In our experiment, the values for $c_{0}$ and $c_{\pi}$ are determined to be 1.03 and 0.73, respectively, from an electron Rabi measurement using standard readout [see Fig.\,\ref{ODMR_Rabi}(b)]. The photon number $n = N_{\text{sweep}} \Bar{n}$ is estimated from a total of $ N_{\text{sweep}}\approx 2 \times 10^{5}$ measurement sweeps (assuming the same time budget as in the SSR experiment), with an average of approximately $\Bar{n}\approx 0.25$ photons collected during the detection window per readout. Consequently, the estimated $\text{SNR}_{\text{std}}$ is 37.

The signal in an SSR-assisted measurement is given by the nuclear spin flip probabilities. Since the C$_{e}$NOT$_{n}$ gate selectively flips the nuclear spin only when the NV electron spin is in the $|0\rangle$ state,  the recorded time trace of normalized counts $I_{\text{norm}}$ after $M$ repetitions of SSR results in a different spin flip probabilities, depending on the electron spin population in $|0\rangle$ state. A typical SSR-assisted nuclear Rabi measurement is illustrated in Fig.\,\ref{ODMR_Rabi}(d), where the maximum spin flip probability is  $f_{0} = 0.70$ and the minimum is $f_{\pi}=0.10$. The resulting flip probability of the nuclear spin $P_{\text{SSR}}$ as a function of the probability $p_0$ to find the NV center in $|0\rangle$ can be expressed as $P_{\text{SSR}} = (f_{0} - f_{\pi})p_0 + f_{\pi}$. The SNR of the SSR-assisted measurement is then given by \cite{haberle2017nuclear}:
\begin{equation}
\text{SNR}_{\text{SSR}}=\frac{|f_{0} - f_{\pi}|\sqrt{M}}{\sqrt{f_{\pi} (1- f_{\pi}) + f_{0} (1 - f_{0})}} ,
\end{equation}
which yields $\text{SNR}_{\text{SSR}}=73$. Therefore, the SNR of the SSR-assisted measurement shows a nearly twofold improvement compared to the standard readout measurement in our experiment.

In the following we evaluate the contributions of the quantum projection noise and photon shot noise in SSR. Specifically, the contributions to the variance of the total number of detected photons in one SSR of the nuclear spin (i.e., $M=1$, see Fig. 2(c) in the main text) due to PSN and QPN, respectively, take the form \cite{Schmitt2021Optimal}:
\begin{equation}
 \sigma^2_{\text{PSN,R}} = (2R)\mu_{|0\rangle} p_0 + (2R)\mu_{|-1\rangle}(1-p_0), \quad 
 \sigma^2_{\text{QPN,R}} = (2R)^2( \mu_{|0\rangle} - \mu_{|-1\rangle} )^2 P_{\text{SSR}}(1-P_{\text{SSR}}) , 
\end{equation}
where $R\approx 700$ is the number of readouts, performed in an alternating manner, leading to the addditional scaling factor of $2$, during one SSR (see Fig. 2(c) in the main text), $\mu_{|0\rangle}=c_{|0\rangle} \bar{n}$ and $\mu_{|-1\rangle}=c_{|-1\rangle} \bar{n}$ denote the average numbers of detected photons per readout for the NV spin states $m_{S}=0$ and $m_{S}=-1$, respectively; $p_0$ represents the population of the $m_{S}=0$ state. One can then calculate the contributions to the variances of the average photon number in one trial SSR, which take the form
\begin{equation}
 \sigma^2_{\text{PSN}} = \frac{\sigma^2_{\text{PSN,R}}}{(2R)^2}= \frac{\mu_{|0\rangle}}{2R} p_0 + \frac{\mu_{|-1\rangle}}{2R}(1-p_0), \quad 
 \sigma^2_{\text{QPN}} = \frac{\sigma^2_{\text{QPN,R}}}{(2R)^2}=( \mu_{|0\rangle} - \mu_{|-1\rangle} )^2 P_{\text{SSR}}(1-P_{\text{SSR}}). 
\end{equation}
Using our experimental parameters $\mu_{|0\rangle}\approx 0.26$, $\mu_{|-1\rangle}\approx0.18$, and $p_0\approx0$ for the case of $\delta_r\approx 0$ and negligible decoherence, we have $\sigma_{\text{PSN}}\approx 0.0113$,  $\sigma_{\text{QPN}}\approx0.0240$. Therefore, quantum projection noise contributes approximately $91\%$ of the total readout noise $\sigma_{\text{total}}\approx(\sigma^2_{\text{PSN}}+\sigma^2_{\text{QPN}})^{1/2}=0.0265$ in this particular example.
These parameters are consistent with earlier reports on ancilla-assisted readout of NV centers \cite{Neumann2010Single, Schmitt2021Optimal}. 

We have demonstrated how the PSN and QPN noise contributions change with SSR with a single NV center, in comparison to the conventional readout method. The analysis shows that QPN is the main noise source with SSR, in contrast to conventional readout, where PSN is typically dominant. By applying the superresolution method, QPN is also suppressed, as shown in the main text. 
Other experiments of quantum sensing with interferometers \cite{Jia2024Squeezing,Feng2025Quantum} and in magnetometry \cite{Troullinou2021Squeezed} typically employ squeezed states of light, in combination with other techniques, e.g., introduction of noise correlations between amplitude and phase noise 
and frequency dependent squeezing, to reduce the effect of the dominant noise source, and surpass the standard quantum limit \cite{Jia2024Squeezing}. However, such and other techniques \cite{Gardner2025prxq} go beyond the scope of our work where we apply laser irradiation for state-dependent readout only, but might be an interesting direction for future experiments. 
Preparing correlated or spin-squeezed states is another route to reduce spin QPN. Theoretical proposals and experimental progress on realizing spin squeezing in solid-state systems, including NV centers, have also been reported \cite{Cappellaro2009Quantum,Bennett2013Phonon,wu2025spinsqueezing}. These typically require a substantial number of NV centers (ensembles) as quantum probes, which is challenging in our current single-quantum probe system and might be an interesting direction for future work.

~\\
\section{Calibration of Signal Amplitudes}

\begin{figure}[t]
\centering
\includegraphics[width=0.6\linewidth]{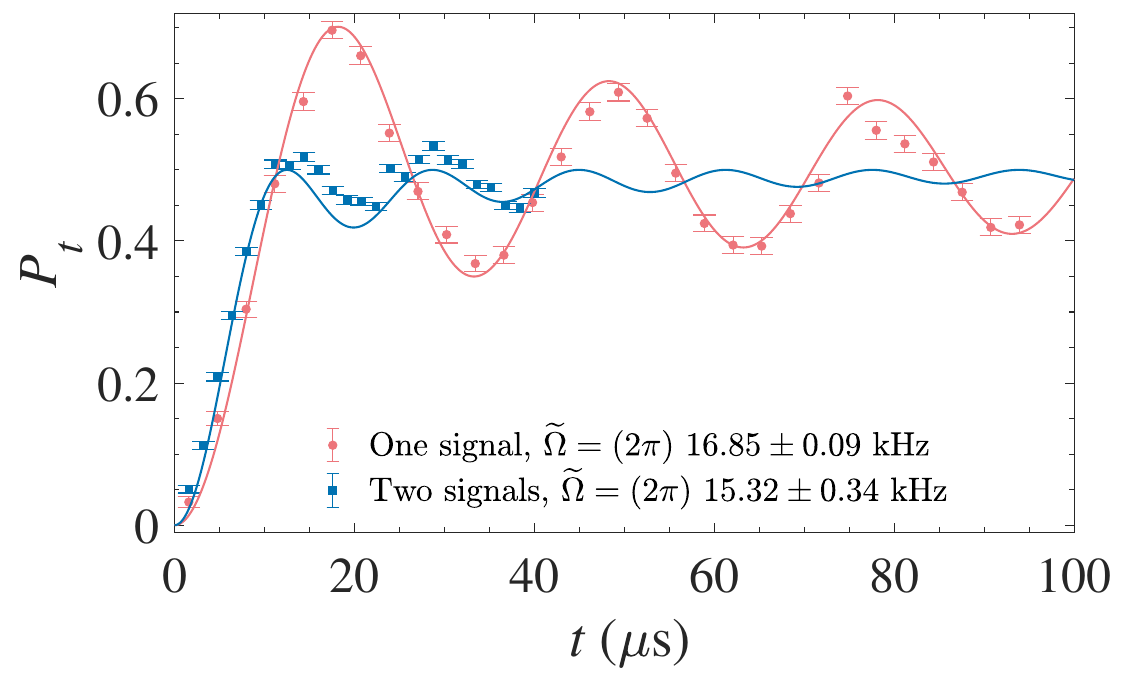}
\caption{ {\bf Calibration of the signal amplitudes by measuring the average transition probability $P_t$.} The red data correspond to the case where a signal with a single frequency of $(2\pi) 2.5125$ MHz was generated from a signal generator. The XY8-$N$ sequence was employed with a fixed interpulse duration of 199 ns, while varying the orders of $N$. The blue data depicts the case where two signals with the frequencies $\omega_{1}=(2\pi) 2.512\;$MHz and $\omega_{2}=(2\pi) 2.513\;$MHz, were applied simultaneously. }
\label{coupling_strength}
\end{figure}

The signal amplitude at the NV center needs to be calibrated prior to every measurement. To refine the estimation of the signal amplitude $\widetilde{\Omega}$ in Eq.\,\eqref{Heff_two_phase}, we first performed an XY8-$N$ experiment to calibrate the signal strength from one of the signal generators. The signal frequency remained consistent at $(2\pi)2.5125$ MHz, and the XY8 pulse frequency was synchronized to this frequency (resulting in an interpulse duration of 199 ns). The number of pulses, $N$, in the XY8 sequence, was varied, and the corresponding measurement results are shown as orange data in Fig.\,\ref{coupling_strength}. The average transition probability, $P_t$, is given by : 
\begin{equation}
\label{Eq:P_amp_one}
P_t=\frac{1}{2}\left[1- J_{0} \left(2\widetilde{\Omega} t\right)\right].
\end{equation}
Through data fitting using Eq.\,\eqref{Eq:P_amp_one}, the signal amplitude was determined to be $\widetilde{\Omega}=(2\pi) 16.85\pm 0.09\,$kHz. Additionally, We performed the same experiment with two signals applied simultaneously, where the frequencies of the two signals were $\omega_{1}=(2\pi)2.512\;$MHz and $\omega_{2}=(2\pi)2.513\;$MHz, respectively. In the presence of both signals, the corresponding $P_t$ is of the form: 
\begin{equation}
\label{Eq:P_amp_two}
P_t= \frac{1}{2}\left[1- J_{0}^{2} \left(2\widetilde{\Omega} t\right)\right].
\end{equation}
The corresponding measurement results are shown as blue data in Fig.\,\ref{coupling_strength}, and by fitting the data to Eq.\,\eqref{Eq:P_amp_two}, the signal amplitude was determined to be $\widetilde{\Omega}=(2\pi) 15.32\pm 0.34\,$kHz.

~\\

\section{Estimation of the uncertainty of the frequency separation}\label{sec:SM_uncertainty_estimation}

In this section, we analyze the uncertainty in the estimation of the frequency separation at the evolution time at $t=80\;\mu$s, corresponding to the superresolution condition. We perform 30 measurements of $C_t$ for every $\delta_r$ to obtain an average, limiting the statistical uncertainty caused by measurement noise, as shown in Fig. 4(b) in the main text. The experimentally measured signal contrast $C_t$ fits excellently the theoretical values. 
Next, we estimate the frequency separation from the measured signal contrast at the superresolution time, 
\begin{equation}
\label{Eq:separation}
\delta_r \approx \frac{\delta_s}{\widetilde{\Omega} t}\sqrt{\frac{1-C_t}{2}}=\frac{2\pi}{\widetilde{\Omega} t^2}\sqrt{\frac{1-C_t}{2}}.
\end{equation}
In Fig. 4(c) in the main text, it can be seen the estimated frequency separation matches well with its actual values set by the experiments for small $\delta_r$, namely, $\delta_r\leq (2\pi) 1$ kHz.

Next, we estimate the corresponding estimation uncertainty. The variance of $\Delta \delta_r$ is given by
\begin{align}
\label{Eq:separation_error}
(\Delta\delta_r)^2 &\approx \underbrace{\left(\frac{\partial\delta_r}{\partial C_t}\Delta C_t\right)^2}_{(\Delta\delta_{r,\Delta C})^2}+\underbrace{\left(\frac{\partial\delta_r}{\partial \delta_s}\Delta \delta_s\right)^2}_{(\Delta\delta_{r,\Delta\delta_s})^2}+\underbrace{\left(\frac{\partial\delta_r}{\partial \widetilde{\Omega}}\Delta \widetilde{\Omega}\right)^2}_{(\Delta\delta_{r,\Delta\widetilde{\Omega}})^2}\notag\\
&=\frac{\delta _s^2}{\widetilde{\Omega }^2 t^2}\frac{(1-C_t) }{2}\left(\frac{\Delta C_t^2}{4
   (1-C_t)^2}+\frac{\Delta \delta_s^2}{\delta _s^2}+\frac{\Delta \widetilde{\Omega }^2}{\widetilde{\Omega }^2}\right),
\end{align}
where $\Delta C_t$ is the experimentally obtained uncertainty in estimating the contrast $C_t$, $\Delta\delta_s$ is the uncertainty in estimating $\delta_s$, and $\Delta \widetilde{\Omega }$ is the uncertainty in estimating $\widetilde{\Omega }$.  We use a Lorentzian function to fit the measured contrast peak for $\delta_r=(2\pi)0$\;kHz in the Fig.\,4(a) of the main text and obtain the estimated peak location $t_{\text{peak}}=79.913 \pm 0.489\;\mu$s. Thus the estimated $\delta_s=(2\pi)12.514\pm 0.077$ kHz.

If we neglect $\Delta\delta_s$ and $\Delta \widetilde{\Omega }$, the uncertainty in estimating the frequency difference takes the form 
\begin{align}
\label{Eq:separation_error_contrast_only}
(\Delta\delta_r)^2 &\approx (\Delta\delta_{r,\Delta C})^2 =\frac{\delta _s^2}{\widetilde{\Omega }^2 t^2}\frac{(1-C_t) }{2}\frac{\Delta C_t^2}{4
   (1-C_t)^2}=\frac{\delta _s^2}{\widetilde{\Omega }^2 t^2}\frac{\Delta C_t^2 }{8(1-C_t)}=\frac{\pi^2}{\widetilde{\Omega}^2 t^4}\frac{\Delta C_t^2}{2(1-C_t)},
\end{align}
where we used that $\delta_s=2\pi/t$ for the last equality. We thus obtain the formula 
\begin{align}
\label{Eq:separation_error_contrast_only_2}
\Delta\delta_{r,\Delta C} &\approx\frac{\delta _s}{2\sqrt{2}\widetilde{\Omega } t}\frac{\Delta C_t }{\sqrt{1-C_t}}= \frac{\pi}{\sqrt{2}\widetilde{\Omega} t^2}\frac{\Delta C_t}{\sqrt{1-C_t}},
\end{align}
as stated in the main text. Table \ref{Ta:std_summarize} summarizes the experimentally measured contrast $C$, along with the corresponding estimated frequency separation $\delta_r$ and its standard deviation $\Delta\delta_{r,\Delta C}$. It should be noted that for $\delta_r = (2\pi)\,$1 and 2.5 kHz, the uncertainties are calculated by directly differentiating the original Bessel function, as the approximate expression Eq.\;\eqref{Eq:separation} is no longer valid in this regime.

\begin{table}[h!]
\centering
\caption{ A summary of experiment results for $\delta_st=2\pi$. Experimental parameters are $\delta_s = (2\pi)$ 12.5 kHz, $t=80\;\mu$s, Repetitions $= 30 \times 4400$. All $\delta_r$ and  $\Delta\delta_{r}$ are in units of $(2\pi)\,$Hz.}
\begin{tabular}{|c|c|c|c|c|c|c|c|c}
\hline
Actual $\delta_r$  & Averaged $C_t$ & $\Delta C_t$ & Estimated $\delta_r$  &  $\Delta\delta_{r,\Delta C}$ &  $\Delta\delta_{r,\Delta\delta_s}$ &  $\Delta\delta_{r,\Delta\widetilde{\Omega}}$ & $\Delta\delta_{r}$\\
\hline
$ 0$     & 0.9998 & 0.0032 & 15.8 & 110.2 & 0.1 & 0.1 & 110.2 \\
\hline
$ 250 $  & 0.9428 & 0.0033 & 252.4 & 7.1  & 1.5 & 1.5 & 7.4 \\
\hline
$ 500 $   & 0.8024 & 0.0034 & 490.0 & 4.0 & 2.9 & 2.8 & 5.6 \\
\hline
$ 1000 $     & 0.3741 & 0.0037 & 970.1 & 4.2 &  22.4 & 5.9 & 23.6 \\
\hline
$ 2500 $   & 0.0555 & 0.0037 & 2583.1 & 30.0 &  90.0  &  21.9 & 97.3 \\
\hline
\end{tabular}\label{Ta:std_summarize}
\end{table}


\section{Limitations and imperfections}\label{sec:SM_imperfections} 

\subsection{Precision, limited by QPN}\label{subsec:SM_imperfections_QPN} 

The analysis presented so far presumes both a perfect signal--probe 
coherence throughout the interrogation time and measurements with 
ideal fidelity. However, in practice, these assumptions are not strictly satisfied. The suppression of projection noise, for instance, relies on the probe state becoming pure when $\omega_r = 0$, which requires the signal to remain coherent over the full measurement interval. If the signal fluctuates within this timescale, complete cancellation of projection noise is no longer possible, thereby imposing a limit. Likewise, finite measurement fidelity produces an effect analogous to signal decoherence, introducing additional noise and setting a practical bound on the achievable resolution.

In this case, $P_t$ approaches zero when $\delta_r\to 0$. In addition, as stated in the main text, $\sigma^2_{\mathrm{QPN}}=P_t(1-P_t)\approx P_{t}$, which implies that quantum projection noise is strongly suppressed when $\delta_r\to 0$, and in practice effectively nullified. Meanwhile, the Fisher information about $\delta_r$ is thus given by:
\begin{equation}\label{Ir_mixed}
	\mathcal{I} (\delta_r)=\frac{(\partial_{\delta_r} P_t)^2}{\sigma^2_{\mathrm{QPN}}}=\frac{(\partial_{\delta_r} P_t)^2}{P_t(1-P_t)}\approx4b_t,
\end{equation}
so it remains constant when $\delta_r\to 0$. Therefore, under the supperresolution condition and with the assumption of perfect measurements, the projection noise is nullified, and a finite $\mathcal{I}(\delta_r)$ is achieved for arbitrarily small $\delta_r$. We note that repeating the experiment $n_{\text{exp}}$ times increases the FI to $n_{\text{exp}}\mathcal{I} (\delta_r)$, where $n_{\text{exp}}=30 \times M$ with $M=4400$ in our particular experiment with $M$ being the number of measurements to obtain a single estimate of the transition probability, which is then repeated 30 times and averaged. Then, the estimation precision is characterised by 
\begin{align}\label{Eq:resolution_no_errors}
(\Delta\delta_r)^2 \ge \frac{1}{n_{\text{exp}}\mathcal{I} (\delta_r)}\approx  \frac{1
   }{4 n_{\text{exp}}b_t}
\end{align}
and should remain constant for arbitrarily low $\delta_r$. 
\\

\subsection{Effect of other noise sources}\label{subsec:SM_imperfections_other}

In realistic experiments, it is typically not possible to completely nullify the quantum projection noise, e.g., due to the presence of other noise sources like fluctuations of the signal during the measurement period, dephasing of the probe, or imperfect measurements. More details on the limitations of the superresolution method are discussed in Ref.~\cite{gefen2019overcoming}. One could usually quantify the resolution limit by including a heuristic noise term, denoted as $\epsilon$, in the transition probability, such that it reads
\begin{equation}\label{p_noise}
	P_{t,\epsilon} = P_{t} + \epsilon \approx b_t\delta_r^2 +  \epsilon,
\end{equation}
This new term imposes a limitation: it is no longer possible to nullify $P_{t,\epsilon}$, which leads to $\mathcal{I}(\delta_r)\to0$ as $\delta_r \to 0$. Under the superresolution condition ($\delta_s t=2\pi$, $\delta_r t\ll 1$, $\epsilon\ll 1$), the Fisher information becomes \cite{gefen2019overcoming} 
\begin{equation}
	\mathcal{I} (\delta_r)\approx\frac{4 b_t^2 \delta _r^2}{(1-2 \epsilon ) b_t \delta _r^2+(1-\epsilon
   ) \epsilon } \approx \frac{4b_{t}^2 \delta_r^2}{b_{t}\delta_r^2+ \epsilon}.
\end{equation}
This expression shows explicitly the dependence of $\mathcal{I}(\delta_r)$ on the additional noise $\epsilon$. 
Therefore, given $\delta_r$, the optimal FI $\mathcal{I} (\delta_r)$ is obviously achieved for $\epsilon = 0$. When $\epsilon\ne0$, the FI $\mathcal{I} (\delta_r)\to 0$ when $\delta_r\to 0$. Thus, the estimation precision is given by 
\begin{align}\label{Eq:resolution_with_errors}
(\Delta\delta_r)^2 \ge \frac{1}{n_{\text{exp}}\mathcal{I} (\delta_r)}\approx \frac{1-2 \epsilon
   }{4 n_{\text{exp}} b_t}+\frac{(1-\epsilon ) \epsilon }{4 n_{\text{exp}}b_t^2 \delta _r^2}\approx \frac{1
   }{4 n_{\text{exp}}b_t}\left(1+\frac{ \epsilon }{ b_t \delta _r^2}\right).
\end{align} 
Therefore, given $\delta_r$, the optimal precision is achieved for $\epsilon = 0$, and it blows up for $\epsilon\ne0$ when $\delta_r\to 0$.

\begin{figure}[t]
\centering
\includegraphics[width=0.55\textwidth]{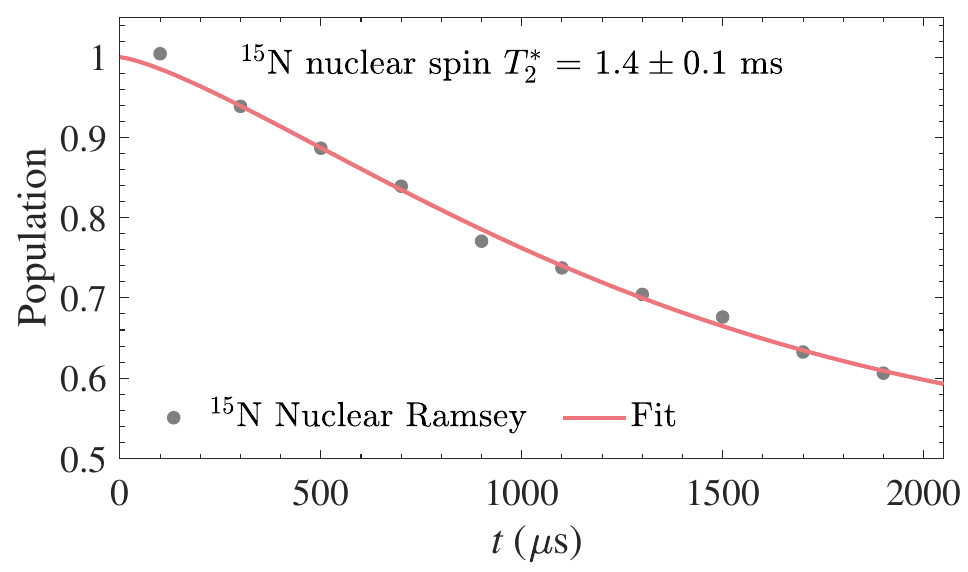}
\caption{The Ramsey experiment of $^{15}$N nuclear spin at a magnetic field $B_{z}=0.6$ T.}
\label{FigR2_N15_ramsey}
\end{figure}

\subsection{Example: decoherence}\label{subsec:SM_imperfections_other}

As a specific example of the above limitation, we note that in actual experiments the measurement time is limited by the coherence time of the quantum sensor, which is  $T_2\approx 1.3$ ms for the electron spin of the NV center in our case. Then, the transition probability can be approximated (for $\delta_r t \ll 1$, $\delta_s t = 2n\pi$) as   
\begin{equation}\label{Eq:P_decoh}
	P_t=\frac{1}{2}\left[1-e^{-\Gamma t} J_0\left(\frac{4 \tilde{\Omega
   } \sin \left(\frac{\delta_r t}{2}\right)}{\delta_r-\delta _s}\right) J_0\left(\frac{4 \tilde{\Omega } \sin
   \left(\frac{\delta_r t}{2}\right)}{\delta_r+\delta
   _s}\right)\right]\approx\frac{1}{2}\left(1- e^{-\Gamma t}\right)+e^{-\Gamma t}\left(b_t\delta_r^2\right),
\end{equation}
where the decay rate $\Gamma\equiv1/T_2$. The corresponding quantum Fisher information takes the form
\begin{equation}\label{Ir_mixed_decay}
	\mathcal{I} (\delta_r)=\frac{(\partial_{\delta_r} P_t)^2}{\sigma^2_{\mathrm{QPN}}}=\frac{(\partial_{\delta_r} P_t)^2}{P_t(1-P_t)}\approx \frac{4 b_t^2 \delta _r^2}{b_t \delta _r^2+(e^{2 \Gamma  t}-1)/4}.
\end{equation}
It is evident that $\mathcal{I} (\delta_r)\to 0$ as $\delta_r\to 0$, so the resolution is limited and determined by Eq. \eqref{Eq:resolution_with_errors} with $\epsilon \approx (e^{2 \Gamma  t}-1)/4$. \\

\subsection{Example: mapping efficiency of the RF pulse for SSR}\label{subsec:SM_imperfections_mapping} 

As another example, we consider the effect of the mapping efficiency of the RF $\pi$ pulse on the protocol. In order to quantitatively assess the performance of the $\pi_{\text{RF}}$ pulse used for mapping the electronic spin state onto the $^{15}$N nuclear spin, we evaluate the average gate fidelity in the presence of pure dephasing. The measured nuclear spin Ramsey dephasing time is  $T_{2,\text{nuclear}}^{\ast} \approx 1.4$ ms, as shown in Fig.\,\ref{FigR2_N15_ramsey}. The $\pi_{\text{RF}}$ pulse duration is $t_\pi \approx 37~\mu\mathrm{s}$. We model the effect of the mapping pulse in a two-level system of states $|m_s=0,m_I=-1/2\rangle$ and $|m_s=0,m_I=+1/2\rangle$ (see Fig. 2(b) in the main text). We ignore any couplings between the nuclear spin states in the $m_s=-1$ manifold as the RF driving field is far detuned as its Rabi frequency $\Omega_n\ll A_{||}$. The system dynamics are modeled by the Hamiltonian in the rotating frame at the RF carrier frequency, which is ideally equal to the Larmor frequency $\omega_n=\gamma_n B_z$ after applying the rotating-wave approximation ($\Omega_n\ll\omega_n$) 
\begin{equation}
 	H_n = \frac{\delta_n}{2}\sigma_z+\frac{\Omega_n}{2}(1+a_n)\sigma_x,
 \end{equation}
where $\delta_n$ is the detuning of the RF frequency from the Larmor frequency, e.g., due to noise, $\Omega_n$ is the Rabi frequency and $a_n$ characterizes the amplitude noise of the RF field. Both $\delta_n$ and $a_n$ can, in general, be time-dependent, but we assume that they are constant during the $\pi$ pulse, which is feasible in our experiment. The time evolution of the system is then described by the propagator $U_n(t)=\exp{(-iH_n t)}$.

The fidelity of the $\pi$ RF pulse can be expressed as 
\begin{equation}
 	F(\Omega_n,\delta_n,a_n) = \left|\frac{1}{2}\text{Tr}(U^\dagger_{n,0}(t) U_n(t))\right|^2=\frac{\Omega _n^2 \left(1+a_n\right)^2}{\Omega
   _{\text{eff}}^2} \sin^2\left(\frac{\Omega _{\text{eff}}}{
   \Omega _n}\frac{\pi}{2}\right),~~\Omega_{\text{eff}}\equiv \sqrt{\Omega_n^2(1+a_n)^2+\delta_n^2},
 \end{equation}
where $t=\pi/\Omega_n=37\,\mu$s and $U_{n,0}(t)$ is the ideal propagator $U_n(t)$ in the absence of errors, i.e., when $\delta_n=0$ and $a_n=0$. The fidelity is also equal to the transition probability of the non-ideal $\pi$ pulse, i.e., the probability that the population of the initial state is transferred to the target state after the pulse. 

We assume that $\delta_n$ is normally distributed with a standard deviation $\sigma_\delta\approx \sqrt{2}/T_{2,\text{nuclear}}^{\ast}\approx(2\pi)\,16$ Hz, which is determined by the dephasing time $T_{2,\text{nuclear}}^{\ast}$ \cite{Salhov2024Protecting}. In addition, the amplitude noise term $a_n$ is also assumed to have a normal distribution with standard deviation $\sigma_a\approx 0.005$, based on previous experiments \cite{Salhov2024Protecting}. The average fidelity is then given by 
\begin{equation}
 	F_{\text{ave}}(\Omega_n,\sigma_\delta,\sigma_a) = \int_{-\infty}^\infty\int_{-\infty}^\infty F(\Omega_n,\delta_n,a_n)p(\delta_n,\sigma_\delta)p(a_n,\sigma_a)d\delta_n da_n\approx 0.9998,
 \end{equation}
where $p(\delta_n,\sigma_\delta)=\frac{1}{\sqrt{2 \pi } \sigma _{\delta }}\exp{\left(-\frac{\delta _n^2}{2 \sigma _{\delta }^2}\right)}$ and $p(a_n,\sigma_a)=\frac{1}{\sqrt{2 \pi } \sigma _{a}}\exp{\left(-\frac{a_n^2}{2 \sigma _{a}^2}\right)}$ are the respective probability density functions. Thus, the average fidelity error $\epsilon_{\text{ave}}\equiv 1-F_{\text{ave}}\approx 0.0002$. The average fidelity is quite high even for a simple $\pi$ pulse and is of the order of the quantum information single qubit fidelity error benchmark of $10^{-4}$. Even for high values of $\delta_n=3\sigma_\delta$ and $a_n=3\sigma_a$, we obtain 
\begin{equation}
F(\Omega_n,\delta_n=3\sigma_\delta,a_n=3\sigma_a)\approx 0.9982,
\end{equation} 
and $\epsilon\approx 0.0018$. We note that the simple $\pi$ fidelity $F=1-\epsilon$ can be improved significantly by using robust pulses, e.g., composite pulses, where $F\approx 1-O(\epsilon^n)$ is feasible \cite{Levitt1986,Genov2014prl}, but this was not necessary in our work. 

Finally, we conclude that, indeed, the fidelity of the RF $\pi$ mapping pulse is negligibly reduced with the average fidelity error of the order of $10^{-4}$. We note that any imperfect mapping efficiency corresponds to an imperfect measurement of the NV spin state. This situation can be described as the transition probability including an additional noise term (denoted as $\epsilon$), which increases the uncertainty of the measurement outcome. Such an effect manifests as an extra contribution to the readout noise and the best achievable resolution, as expressed in Eq.\eqref{Eq:resolution_with_errors}, and is consistent with the discussion provided.

~\\


\section{Correspondence of the experimental data to the theoretical Quantum Fisher Information}\label{sec:SM_Experiment_QFI}

\begin{figure}[t]
\centering
\includegraphics[width=0.6\textwidth]{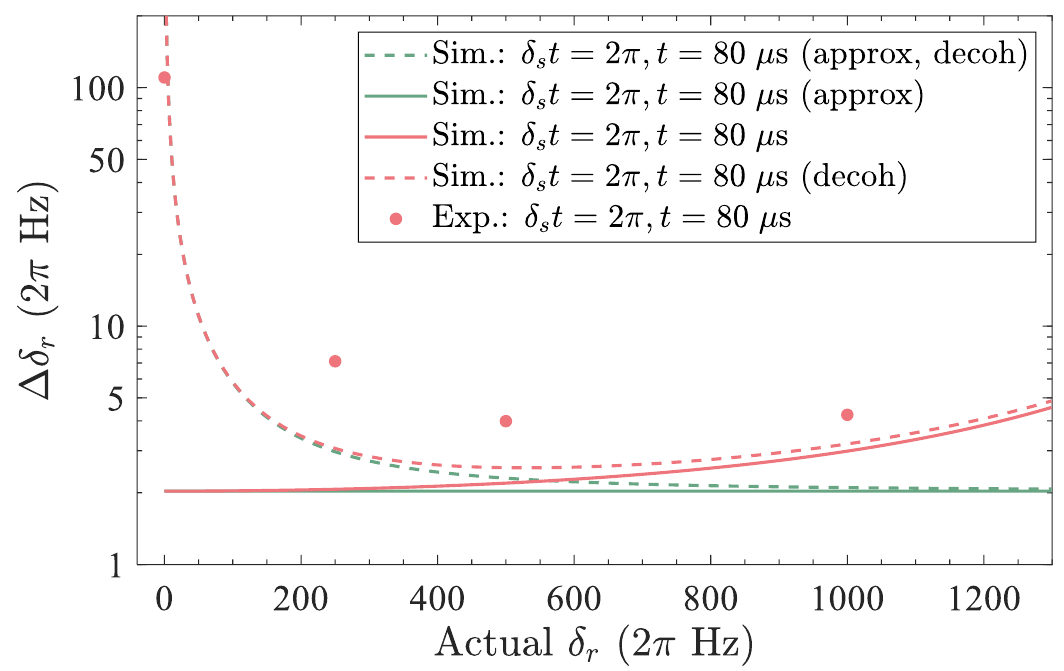}
\caption{ Estimation uncertainty $\Delta\delta_r$ versus $\delta_r$. The red dots represent the experimental results measured under superresolution condition $\delta_s t= 2\pi$ with $\delta_s = (2\pi)\,12.5$ kHz and $t = 80\,\mu$s. The red solid line shows the theoretical limit from Fisher information under the assumption of perfect measurements, calculated using Eq.\,\eqref{Eq:QFI-bound} and Eq.\,\eqref{pa_phase} with the same experimental parameters. The red dashed line represents the simulated $\Delta\delta_r$ obtained including the decoherence of the quantum sensor, as described by Eq.\,\eqref{Eq:P_decoh} and Eq.\,\eqref{Ir_mixed_decay}, without approximation. This plot is extracted from Fig.\,4(d) for clearer comparison. Additionally, the green dashed and solid lines show the corresponding theoretical limits obtained from the approximated Fisher information (Eq.\,\eqref{Ir_mixed_decay} and Eq.\,\eqref{Ir_mixed}), with and without incorporating sensor decoherence, respectively. For $\delta_r > (2\pi)1$ kHz the red and green lines systematically diverge, indicating that the approximations leading to Eq.\,\eqref{Ir_mixed} break down beyond $\delta_r\approx(2\pi)\,1$ kHz for our experimental parameters. This is consistent with the conclusion presented in Fig.\;\ref{P_difference}.
}
\label{FigR3_freq_std}
\end{figure}

In order to compare the experimental data with the theoretical values of QFI, we perform simulations of the expected precision and compare with experimental results. Specifically, we use the formula in Eq. \eqref{Ir_mixed_decay} with the formula for the expression $P_t$ taken from the exact expression in Eq. \eqref{Eq:P_decoh} with the respective Bessel functions, i.e., without the approximation for small $\delta_r$, for the simulation of the precision in Fig. 4(d) in the main text. 
The parameters $\delta_s$ and $t$ are chosen according to the superresolution condition, $\delta_s t = 2\pi$. Using the experimental values of $\{\delta_s, \Omega, t\}$, we simulate the optimal $\Delta\delta_r$ predicted by the QFI. This result for $\Gamma=0$ (without decoherence) is plotted as the red solid line in Fig.\,4(d) of the main text, in direct comparison with the experimental data (red dots). In addition, to make the comparison clearer, we have extracted only the experimental data and the theoretical limit set by the Fisher information from Fig.\,4(d) and replotted them separately in Fig.\,\ref{FigR3_freq_std}. Similarly to Fig.\,4(d), Fig.\,\ref{FigR3_freq_std} shows the estimation uncertainty $\Delta \delta_r$ as a function of $\delta_r$: the red dots represent the values of $\Delta \delta_r$ obtained experimentally under the superresolution condition with $\delta_s = (2\pi)\,12.5$ kHz and $t = 80\,\mu$s, while the red solid line represents the ideal values of $\Delta \delta_r$ calculated from the quantum Fisher information using the same experimental parameters. 

As shown, the agreement between experiment and theory is very good when $\delta_r$ lies between $(2\pi)\,250$ Hz and $(2\pi)\,1$ kHz. 
For smaller values of $\delta_r$ (below $(2\pi)\,250$ Hz), a clear deviation  emerges from the case when quantum projection noise (QPN) is dominant. Theoretically, because QPN is strongly suppressed, the Fisher information remains finite for arbitrarily low $\delta_r$, predicting that $\Delta\delta_r$ should remain relatively constant for very small $\delta_r$. Experimentally, however, $\Delta\delta_r$ increases as $\delta_r$ becomes smaller. This discrepancy arises because additional noise sources---such as decoherence and residual classical readout noise---contribute a nearly constant background noise term, which dominates in this regime, as discussed in Sections \ref{sec:SM_uncertainty_estimation} and \ref{sec:SM_imperfections}. This is confirmed from the simulations for $\Gamma=1/T_2,~T_2\approx 1.3$ ms (red, dashed line) in Fig. 4(d) in the main text and Fig.\,\ref{FigR3_freq_std}, which use Eq. \eqref{Ir_mixed_decay}, accounting for decoherence. The simulation with decoherence fits the experimental data very well even for small $\delta_r$.  The experimental uncertainty is overall slightly higher than the theoretical expectation most likely due to the presence of additional noise and errors, which are not accounted to the simulations, e.g., classical readout noise, SSR infidelity, and mapping errors. In addition, based on this simulation and the resolution criterion $\Delta\delta_r \leq |\delta_r|$, the best achievable frequency resolution under our experimental conditions is estimated to be $(2\pi)23.3$ Hz.


\end{document}